\documentclass[referee]{raa}            

\usepackage{graphicx,times}             
\usepackage{natbib}
\usepackage{amssymb,amsmath}
\bibpunct{(}{)}{;}{a}{}{,}

\usepackage[pagebackref=true]{hyperref}

\usepackage{longtable}
\usepackage{multirow}
\usepackage{threeparttablex}
\usepackage{booktabs}
\usepackage{xcolor}

\newcommand{\LCDM}{$\Lambda$CDM}
\newcommand{\HI}{H\,I}
\newcommand{\kms}{km\,s$^{-1}$}

\begin{document}

  \title{Optical identification of the FASHI sources: toward the extended Local Volume}

   \volnopage{Vol.0 (20xx) No.0, 000--000}      
   \setcounter{page}{1}          

   \author{Aleksandra E. Nazarova \inst{1}
   \and Dmitry I. Makarov \inst{1}
   \and Igor D. Karachentsev \inst{1}
   \and Chuan-Peng Zhang \inst{2,3}
   \and Maksim I. Chazov \inst{1}
   \and Ming~Zhu \inst{2,3}}

   \institute{Special Astrophysical Observatory, Russian Academy of Sciences, Nizhnii Arkhyz, 369167 Russia; {\it a.e.nazarova@sao.ru}\\
   \and 
   National Astronomical Observatories, Chinese Academy of Sciences, Beijing, 1000101, PR China\\
   \and 
   Guizhou Radio Astronomical Observatory, Guizhou University, Guiyang 550000, PR China\\
\vs\no
   {\small Received 20xx month day; accepted 20xx month day}}

\abstract{ We extracted a list of 662 nearby (within $\sim16$ Mpc) HI-detection sources from the Five-hundred-meter Aperture Spherical radio Telescope (FAST) All Sky HI Survey (FASHI) and made a visual identification of them with optical counterpart. This inspection led to the discovery of 71 new dwarf galaxies. All of them are dwarf irregular galaxies with ongoing star formation. 
They are characterized by the following median parameters: visual magnitude of $g=17.8$~mag and color $(g-r)=0.29$~mag, \HI{}-flux  $S_\mathrm{HI}=718$~mJy\,\kms{}, \HI{}-mass $M_\mathrm{HI}=3.7\times10^7$~$M_\odot$, as well as the \HI{} line-width of $W_{50}=37$~\kms{}.
}

   \authorrunning{Nazarova et al.}
   \titlerunning{Optical identification of the FASHI sources: toward the extended Local Volume}

   \maketitle
   
\section{Introduction}           
\label{sect:intro}

The Local Volume (LV), conventionally limited to a radius of 10--11~Mpc, is one of the most important laboratories for studying the Universe~\citep{2004AJ....127.2031K, 2013AJ....145..101K}. 
The proximity of LV galaxies allows us to see their resolved stellar populations and reconstruct star formation and evolution histories~\citep[see for instance,][]{2010MNRAS.406.1152M, 2011ApJ...739....5W, 2014ApJ...789..147W, 2017MNRAS.466..556M, 2021MNRAS.502.1623M}. 
The LV is the basis of the extragalactic distance scale~\citep{2001ApJ...553...47F, 2007ApJ...661..815R, 2024ApJ...966...89A}. 
The Hubble Space Telescope provides massive measurements of high-precision distances using the TRGB method \citep{2009AJ....138..332J, 2021AJ....162...80A} to map the detailed distribution of galaxies, measure the peculiar velocity field and detect cosmic flows, and study the distribution of dark matter~\citep{1996AJ....111..794K, 2008ApJ...676..184T, 2009MNRAS.393.1265K, 2012AstBu..67..123K, 2014ApJ...782....4K, 2015ApJ...805..144K, 2025A&A...698A.178M}. 
It is important to note that many contradictions with the standard \LCDM{} model appear on small scales and can only be detected in the nearest Universe~\citep{2004ApJ...609..482K, 2014MNRAS.444..222G, 2015MNRAS.454.1798K, 2017Galax...5...17D}.

Most sky surveys are limited by detected flux or apparent magnitude.
This results in a predominance of giant galaxies due to the larger effective survey depth for bright objects and, consequently, a larger volume surveyed.
This leads to an incomplete picture for the faint end of the galaxy population, where low-mass galaxies reside.
In contrast, the LV is limited by distance.
Although forming such a sample proves to be a more difficult task, this approach provides a more homogeneous representation of different populations of galaxies.
Due to its proximity, LV allows us to study in detail the population of dwarf galaxies that become inaccessible at greater distances. 
According to the database of LV galaxies~\citep{2012AstBu..67..115K}, dwarf galaxies with masses $<10^9$~$M_\odot$ represent 85\% of the LV population~\citep{2019AstBu..74..111K}. 
This fact allows us to study the low-mass tail of the luminosity and mass functions with high accuracy.
The luminosity function indicates that the LV is complete at least down to $M_B=-14$~mag~\citep{2015MNRAS.454.1798K}.
Advancing into the faint luminosity range requires a sustained effort to find new dwarf galaxies.
The systematic search for galaxies has significantly improved the completeness of the data~\cite[a short and incomplete list of recent studies is][]{2020AN....341.1037K, 2022AstBu..77..372K, 2023MNRAS.521..840K, 2023A&A...678A..16K, 2024Ap.....66..441K}.

Blind 21-cm sky surveys such as the \HI{} Parkes All Sky Survey~\citep[HIPASS,][]{1997ApJ...490..173Z}, the Arecibo Legacy Fast ALFA~\citep[ALFALFA,][]{2005AJ....130.2598G}, FAST All Sky \HI{}~\citep[FASHI,][]{2024SCPMA..6719511Z} play a special role in studying the nearby Universe. 
They allow us to study the \HI{} mass function and gas reservoirs of galaxies; 
to detect galaxies with unusual properties, such as Leo~P~\citep{2013AJ....146...15G}, a galaxy with a large \HI{} reservoir;
to construct a circular velocity function~\citep{2011ApJ...739...38P, 2015MNRAS.454.1798K} that is substantially simpler to compare with theory predictions than the luminosity function; 
and to discover new galaxies of low surface brightness (LSB).
In a recent paper \citet{2024A&A...684L..24K} found 20 new LV galaxies using FASHI data~\citep{2024SCPMA..6719511Z}. 

Despite the importance of the LV for studying the Universe, it has one major drawback --- it covers a rather small region of space.
On scales of the order of 10 Mpc, the relative matter density fluctuations are rather large of the order of $\sigma_8\approx0.81$~\citep{2020A&A...641A...1P}.
To minimize this effect, it is very important to increase the depth of the distance-limited sample.

In this work, we extended the search for nearby galaxies from the FASHI survey to a distance of about 16~Mpc and performed an optical identification of all FASHI sources using the DESI Legacy Imaging Surveys~\citep{2019AJ....157..168D}, Pan-STARRS~\citep{2016arXiv161205560C}, SDSS~\citep{2000AJ....120.1579Y} and other optical sky surveys.

\section{Optical identification}
\label{sec:inspection}

The FASHI survey~\citep{2024SCPMA..6719511Z} plans to scan the entire sky in the declination range between $-14^\circ$ and $+66^\circ$ in the 1050 to 1450~MHz range using a 19-beam receiver of the Five-hundred-meter Aperture Spherical radio Telescope (FAST).
The median detection sensitivity is $\sim 0.76$~mJy\,beam$^{-1}$ at a velocity resolution of 6.4~\kms{} after smoothing.
The spatial resolution of the FAST is $2\farcm9$ at 1420~MHz.
The first release of the FAST survey covers the region above and below the ALF\-ALFA survey~\citep{2005AJ....130.2598G}, namely in the range of declination of $[-6^\circ,0^\circ]$ and $[+30^\circ,+66^\circ]$ and in the range of right ascensions of $[0^\mathrm{h},17.3^\mathrm{h}]$ and $[22^\mathrm{h},24^\mathrm{h}]$ (see Figure~\ref{fig:map}).
The catalog contains 41741 extragalactic \HI{} sources in the heliocentric redshift range $200<cz_\odot<26323$~\kms{}.
67\% of them were cross-identified with the Siena Galaxy Atlas~\citep[SGA,][]{2023ApJS..269....3M} and the SDSS DR7 catalog~\citep{2009ApJS..182..543A}.

In line with our goal of expanding the sample of LV galaxies, we extracted a list of 662 FASHI objects with velocities $V_\mathrm{LG}\leq1150$~\kms{} in the system of the Local Group centroid~\citep{1996AJ....111..794K}, which roughly corresponds to 16~Mpc.
Many of them have already been identified with known galaxies as part of the FASHI survey (111 with ALFALFA, 381 with SGA, 34 and 156 with SDSS with and without known redshifts); and 178 of them correspond to well-known LV galaxies.
However, a substantial portion of the sources (42\% in comparison with SGA sample) turned out to be without optical identification.
We decided not to limit ourselves to unidentified sources, but to perform a full visual inspection of all 662 objects.
For this purpose, we used the Interactive Map\footnote{\url{https://www.legacysurvey.org/viewer}} of the 10th release of the DESI Legacy Imaging surveys~\citep{2019AJ....157..168D}.
In cases where the study area was not covered by the DESI Legacy surveys or the object could not be identified, we used images from other surveys, such as SDSS, Pan-STARRS, unWISE, and DSS, using the Aladin Sky Atlas\footnote{\url{https://aladin.cds.unistra.fr/}}~\citep{2000A&AS..143...33B}.
During identification, we took into account the radial velocity of the galaxy, which should be within $\pm 30\div50$ km/s from radial velocity of the \HI{}-detection source; proximity of the coordinates of the \HI{}-detection source to the possible optical counterpart, which should lie within $2.9\arcmin$ (beam size at 1420~MHz); the morphology of the galaxy, which should be of a late type for gas-rich systems; as well as total flux and \HI{} line-width, which should vary during the transition from dwarf to massive galaxies.
Due to the relatively high resolution of the FASHI survey and the proximity of the LV galaxies, the problem of ambiguity did not arise, and the HI-detection sources were linked to their optical counterparts with high confidence.
Further identification was performed using the LV galaxies database\footnote{\url{http://www.sao.ru/lv/lvgdb/}}~\citep[LVGDB,][]{2012AstBu..67..115K}, HyperLeda\footnote{\url{https://leda.univ-lyon1.fr/}}~\citep{2014A&A...570A..13M}, and NASA/IPAC Extragalactic Database\footnote{\url{https://ned.ipac.caltech.edu/}}~\citep[NED,][]{1991ASSL..171...89H}.
As a result, we created a list of cross-identifications of FASHI sources with known galaxies, discoveries of new galaxies, and empty fields without a suitable optical counterpart.

\setlength{\tabcolsep}{1pt}
\begin{table}
\caption{Example of the identification table for 662 FASHI objects}\label{tab:list_all}
\medskip
\begin{tabular}{c c c c c c c c c c c}
\hline\hline
N &
FASHI ID &
J2000$_\mathrm{FASHI}$ & 
J2000$_\mathrm{opt}$ &
MainName &
PGC &
$V_\mathrm{LG}$ &
\scriptsize{ALFALFA} &
\scriptsize{SGA} &
\scriptsize{SDSS$_\mathrm{s}$} &
\scriptsize{SDSS$_\mathrm{ph}$} \\
(1) & (2) & (3) & (4) & (5) & (6) & (7) & (8) & (9) & (10) & (11) \\
\hline
2   & 20230006805 & J120912.10+305418.5   & J120911.81+305424.7   & UGC07131     & 038598 & 226 & + &  +  &   &     \\
18  & 20230009357 & J123838.55+324552.1   & J123840.01+324601.2   & UGCA292      & 042275 & 304 & + &     &   & $-$ \\
102 & 20230001478 & J060406.77$-$043738.1 & J060407.71$-$043824.7 &\textit{new}  &        & 532 &   &     &   &     \\
338 & 20230013075 & J112033.04+355404.9   &                       &\textit{empty}&        & 808 &   &     &   & $-$ \\
510 & 20230003668 & J130058.56$-$020115.1 & J130059.63$-$020107.9 & PGC135815    & 135815 & 986 &   & $-$ &   &     \\ 
\hline\hline
\end{tabular}
\end{table}

Table~\ref{tab:list_all} shows an example of identifications (the entire table is available only in electronic form). 
The columns contain the following information:
(1) row number, table sorted by increasing $V_\mathrm{LG}$ (see column (7));
(2) ID number of the radio source from the FASHI catalog;
(3) the J2000.0 equatorial coordinates of the \HI{}-source;
(4) the J2000.0 equatorial coordinates of the optical counterpart, if exist;
(5) main name of the optical counterpart, if exist (\textit{new} and \textit{empty} mark sources from corresponding lists in current study);
(6) PGC-number from the HyperLeda database, if exist;
(7) radial velocity of the \HI{}-source relative to the Local Group centroid (\kms);
(8)--(11) information on the existence and correctness of automatic FASHI identification (``+'' marks correct and ``$-$'' indicates incorrect identifications).

We confirm all automatic FASHI identifications with ALF\-ALFA and SGA, except for three cases:
\begin{description}

\item[ 20230029833 ] 
$cz_\odot=744$~\kms{} coincides with a NGC\,3769. However, FASHI identified it with its companion NGC\,3769A. We believe that ID$_\mathrm{FASHI} = 20230029833$ is associated with NGC\,3769, the main galaxy in the interacting pair.

\item[ 20230061084 ] 
$cz_\odot=607$~\kms{}. FASHI identified it with PGC\,100707, but its known \HI{} width $W_{50}=27$~\kms{} and velocity $cz_\odot=692$~\kms{} are inconsistent with FASHI data $W_{50}=198$~\kms{} and $cz_\odot=607$~\kms{}. In fact, this signal comes from the disturbed \HI{} distribution in the NGC\,4631 + 4656 group. Thus, we decided to classify it as a source without optical identification (see Table~\ref{tab:list_empty}).

\item[ 20230003668 ] 
$cz_\odot=1131$~\kms{} best matches PGC\,135815 in position and morphology instead of PGC\,214054 in the FASHI SGA identification.
The galaxy PGC\,214054 is $2\farcm6$ North of the FASHI coordinates, and more importantly, it has an optical SDSS velocity measurement, $cz_\odot=1380.2\pm1.9$~\kms{}, which contradicts the FASHI data.

\end{description}

In addition, we found a problem with the identification of ID$_\mathrm{FASHI} = 20230029002$, which $V_\mathrm{LG} = 1184$~\kms{} is slightly outside of our sample limit of $V_\mathrm{LG} = 1150$~\kms{} and therefore not included in Table~\ref{tab:list_all}.

\begin{description}
\item[ 20230029002 ] 
$cz_\odot=1126$~\kms{} corresponds to SDSS\,J121759.40+465634.3 ($cz_\odot=1149\pm10$~\kms{}).
The identification of FASHI SGA with SDSS\,J121811.07+465500.6 = [KKH2011]\,S05 = LV\,J1218+4655 at $2\farcm5$ SE of the FASHI coordinates looks wrong.
LV\,J1218+4655 is a well known satellite of NGC\,4258 with a TRGB distance of $D=8.2$~Mpc and with an optical SDSS velocity of $cz_\odot=402.1\pm1.5$~\kms{}.
\end{description}

Unfortunately, the FASHI identification with SDSS is not so successful, mainly in the case of SDSS objects without redshifts. 
Given the morphology of the galaxies and the presence of better optical counterparts, we reject 92 of 190 FASHI identifications with SDSS (only 2 with known SDSS redshift), which is 48\% of the sample. 

\begin{figure*}
\centering
\includegraphics[width=\textwidth]{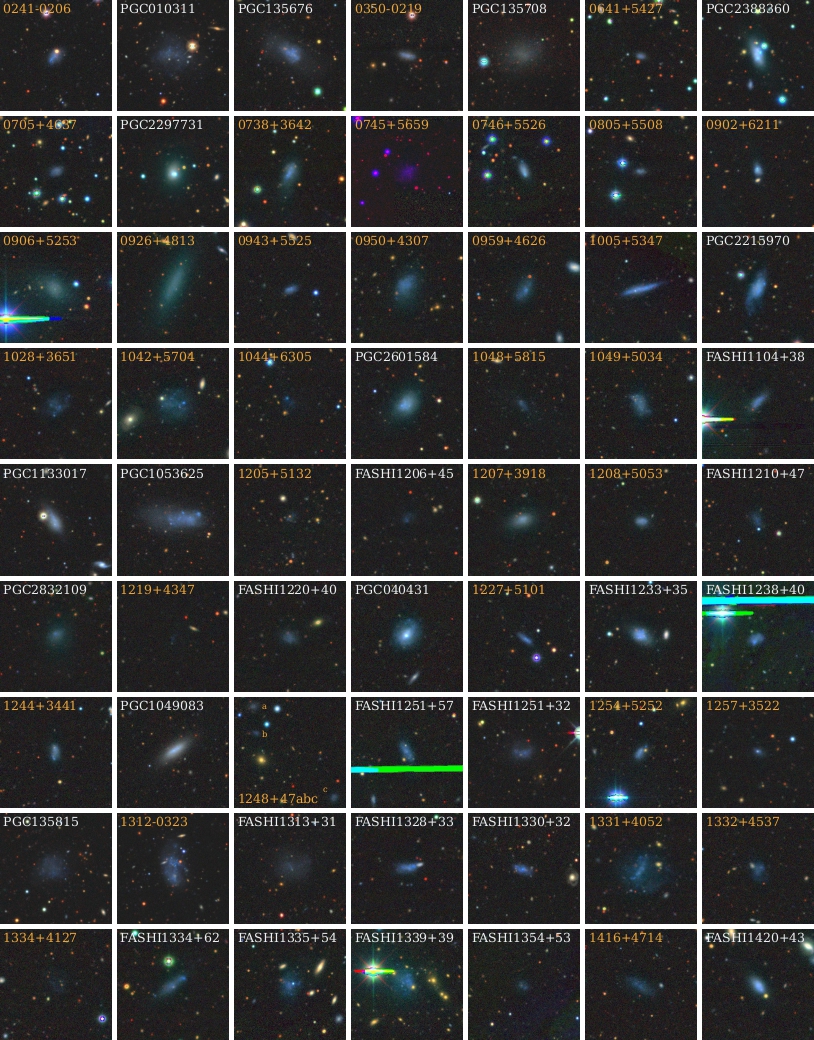}
\includegraphics[width=\textwidth]{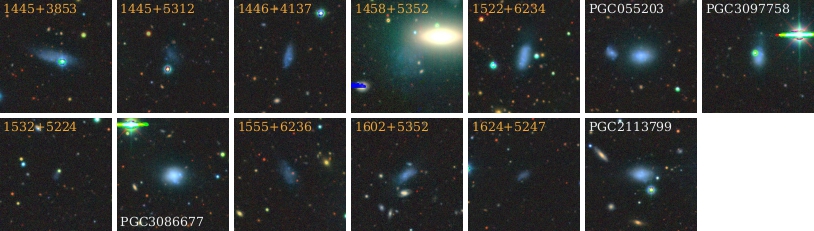}
\caption{Images of the new Local Volume candidate galaxies taken from DESI Legacy Imaging Surveys (DR-10). Each panel   is~$2\arcmin\times2\arcmin$, North is to the top and East is to the left. An orange label indicates a new galaxy and white label indicates a previously identified system. Some galaxies are not in the DR-10 footprint and not included in the image.}
\label{fig:mosaic}
\end{figure*}


The following are comments on some of the objects.
\begin{description}
\item[ 20230026779 ] 
$cz_\odot=713$~\kms{} fits a dwarf galaxy UGC\,672 = PGC\,3905. However, the SDSS mistakenly identifies a foreground star as the galaxy center.

\item[ 20230022506 ] 
$cz_\odot=521$~\kms{} corresponds to a dwarf galaxy PGC\,3096870 with an optical $cz_\odot=523 \pm 82$~\kms, which is in excellent agreement with FASHI data.
However, the SDSS wrongly points to a distant background galaxy as its center.

\item[ 20230036966 ] 
$cz_\odot=754$~\kms{} coincides with a famous metal-poor galaxy, I\,Zw\,18. However, FASHI identified it with I\,Zw\,18B component, which is usually interpreted as either a companion dwarf or a star-forming clump. We believe that ID$_\mathrm{FASHI} = 20230036966$ is associated with I\,Zw\,18A, the main body of the galaxy.

\item[ 20230024488 ] 
$cz_\odot=843$~\kms{} is identified in FASHI with SDSS\,J120712.31+425438.8, but its morphology and magnitude $g=21.54$ are inconsistent with \HI{}-parameters $W_{50}=215$~\kms{} and $S_\mathrm{HI}=11.5$~Jy\,\kms{}. In fact, this signal is from the \HI{} distribution in a group of galaxies around NGC\,4111 without optical identification (see Table~\ref{tab:list_empty}).

\item[ 20230114123 ] 
$cz_\odot=1059$~\kms{} is associated in FASHI with a galaxy SDSS\,J131109.93+363043.5, but this is a distant galaxy and cannot be its optical counterpart. This signal appears to be the result of the interaction between NGC\,5033 and NGC\,5002 in the NGC\,5005 group.

\item[ 20230057619 ] 
$cz_\odot=825$~\kms{} corresponds to an uncatalogued dwarf galaxy J155532.59+623618.0. FASHI linked it to a distant galaxy SDSS\,J155531.97+623619.2, which lies at the edge of the dwarf galaxy, but SDSS photometry includes the entire region of the dwarf galaxy.

\end{description}

\subsection{New Local Volume candidates}

We discovered 71 new faint dwarf galaxies $16.4 \lesssim g \lesssim 19.2$~mag that were previously uncatalogued.
Their $2\arcmin\times2\arcmin$ images (North to the up and East to the left) from the DESI Legacy Imaging Surveys~\citep{2019AJ....157..168D} are presented in Fig.~\ref{fig:mosaic}.
All objects are irregular type galaxies.
They are characterized by a typical median color of $(g-r)=0.29$~mag and median \HI-flux of $S_\mathrm{HI}=718$~mJy\,\kms{}.

Two dwarfs, ID$_\mathrm{FASHI} = 20230063962$ $cz_{\sun}=266$~\kms{} and ID$_\mathrm{FASHI} = 20230001478$ $cz_{\sun}=675$~\kms{}, fall within the LV bounds $V_\mathrm{LG}\leq550$~\kms{}.
This gives a significant gain to the 20 new LV galaxies recently discovered by~\citet{2024A&A...684L..24K}.
\begin{description}
\item[ 20230007694 ] $cz_\odot=720$ is hidden by the Dark cloud LDN\,1523 = DOBASHI\,4393 with extremely large extinction and is only visible in IR in the WISE survey images.
\end{description}

The basic parameters of these galaxies are summarized in Table~\ref{tab:list_new} (see Appendix~\ref{app:ListNew}).
In addition to the new discovered objects, the table includes galaxies for which the line-of-sight velocities were not previously known.
For these galaxies we provide a common name.
The table also contains 20 objects found previously \citet{2024A&A...684L..24K}, which have names like FASHI0237+38, etc.
\HI{}-profiles of the new LV candidate galaxies from FASHI are given in the same sequence as in Table~\ref{tab:list_new} (see Appendix~\ref{app:Spectra}).

Figure~\ref{fig:map} shows the sky distribution of \HI{} sources from our sample associated with known galaxies (blue), along with newly discovered galaxies (red). 
The distribution is supplemented by known nearby galaxies from the LV galaxy database. 
As expected, the identified galaxies trace the distribution of LV galaxies with a clear concentration along the plane of the Local Supercluster. 
In contrast, the newly discovered galaxies are distributed relatively uniformly across the entire FASHI coverage area. 
Note that several of the new galaxies are located at low Galactic latitudes.

\begin{figure}
\centering
\includegraphics[width=\textwidth]{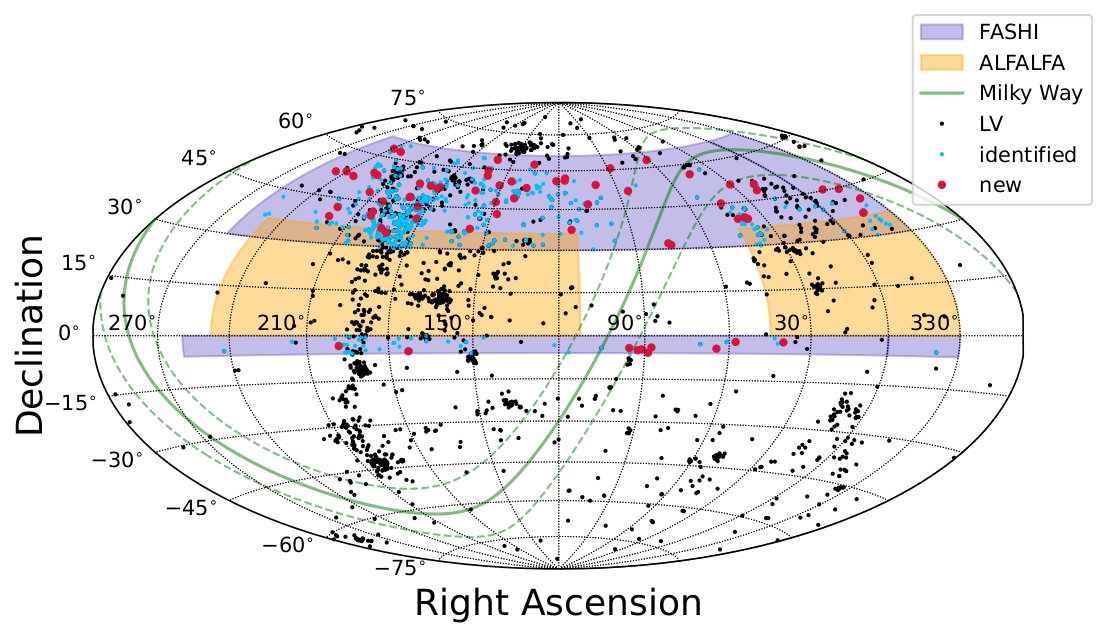}
\caption{
Footprints of FASHI and ALFALFA surveys with distribution of previously known LV galaxies (black), new galaxies from our sample (red) and other identified galaxies from our sample (blue) on the sky in equatorial coordinates. The zone of avoidance in the Milky Way is outlined by the green line.
}
\label{fig:map}
\end{figure}

For 62 of the new LV candidate galaxies, we performed surface brightness photometry in the $g$ and $r$ bands using the DESI Legacy Imaging Surveys images. The results are presented in Table~\ref{tab:photometry} in Appendix~\ref{app:Photometry}, which contains data in both filters on the asymptotic total magnitude $m$, the effective radius $r_{50}$, and the radius $r_{80}$, containing 50 and 80\% of the total flux, respectively. In several cases, the corresponding values were taken from the SGA~\citep{2023ApJS..269....3M}. The table also shows the total apparent magnitudes $B$ and $V$ estimated from the $g$ magnitude and $(g - r)$ color using the transformation equations from the Dark Energy Survey~\citep{2021ApJS..255...20A} and the $m_{21}$ magnitude estimated as $m_{21}= 17.4 - 2.5 \log(S_\mathrm{HI})$ which can be used for the gas-dominated galaxies instead of the $B$-magnitude for the Baryon Tully-Fisher relation~\citep{2025AJ....170...23N}. To estimate the stellar mass (listed in Table~\ref{tab:list_new}), we use the $M/L$ relation $\log(M_*/L_B) = -0.91 + 1.45(B - V)$ justified by \citet{herr2016}, where $B$ and $V$ magnitudes are corrected for the Galactic extinction.

\subsection{Empty fields}


Our sample contains 49 \HI{} sources that could not be reliably associated with any optical counterpart. These cases are presented in Table~\ref{tab:list_empty} (see Appendix~\ref{app:Empty}). 

As one can see, some of these sources are close to the plane of the Galaxy having a low Galactic latitude, which makes the detection of an optical counterpart extremely unlikely.
Also, some sources are associated with well-known nearby galaxy groups falling within their virial zones, and thus may represent a detection of complex gas structure in the system during the interaction.
In the case of sources with large line-widths, such as ID$_\mathrm{FASHI}=20230057471$, 20230025619, and 20230001598, the presence of an object similar to the Large Magellanic Cloud is expected, which would be impossible to miss in modern optical sky surveys. We conclude that these signatures may be the result of data processing issues.
Another object ID$_\mathrm{FASHI}=20230004629$ with an unusually narrow line $W_\mathrm{50} = 7$~\kms{} also looks like a processing artifact.
We checked a dozen empty fields and confirmed the false positive detection of the \HI{} signal. 
These fields are marked as wrong detection in Table~\ref{tab:list_empty}.
Thus, after excluding the cases described above, the only one \HI{}-source ID$_\mathrm{FASHI} = 20230057481$ with $S_\mathrm{HI}=0.55$~Jy\,\kms{} and $W_{50} = 43$~\kms{} left, which fits a dwarf galaxy and does not have an optical counterpart and may be a dark galaxy with extremely low surface brightness.
Unfortunately, this object is also located at a low Galactic latitude of $b\approx-15.2$, where the absorption reaches 2.3~mag in the $B$-band, which significantly limits the possibility of detecting its optical counterpart.

\subsection{Velocity confirmation for PGC\,168301}

\begin{figure}
\centering
\includegraphics[width=0.8\linewidth]{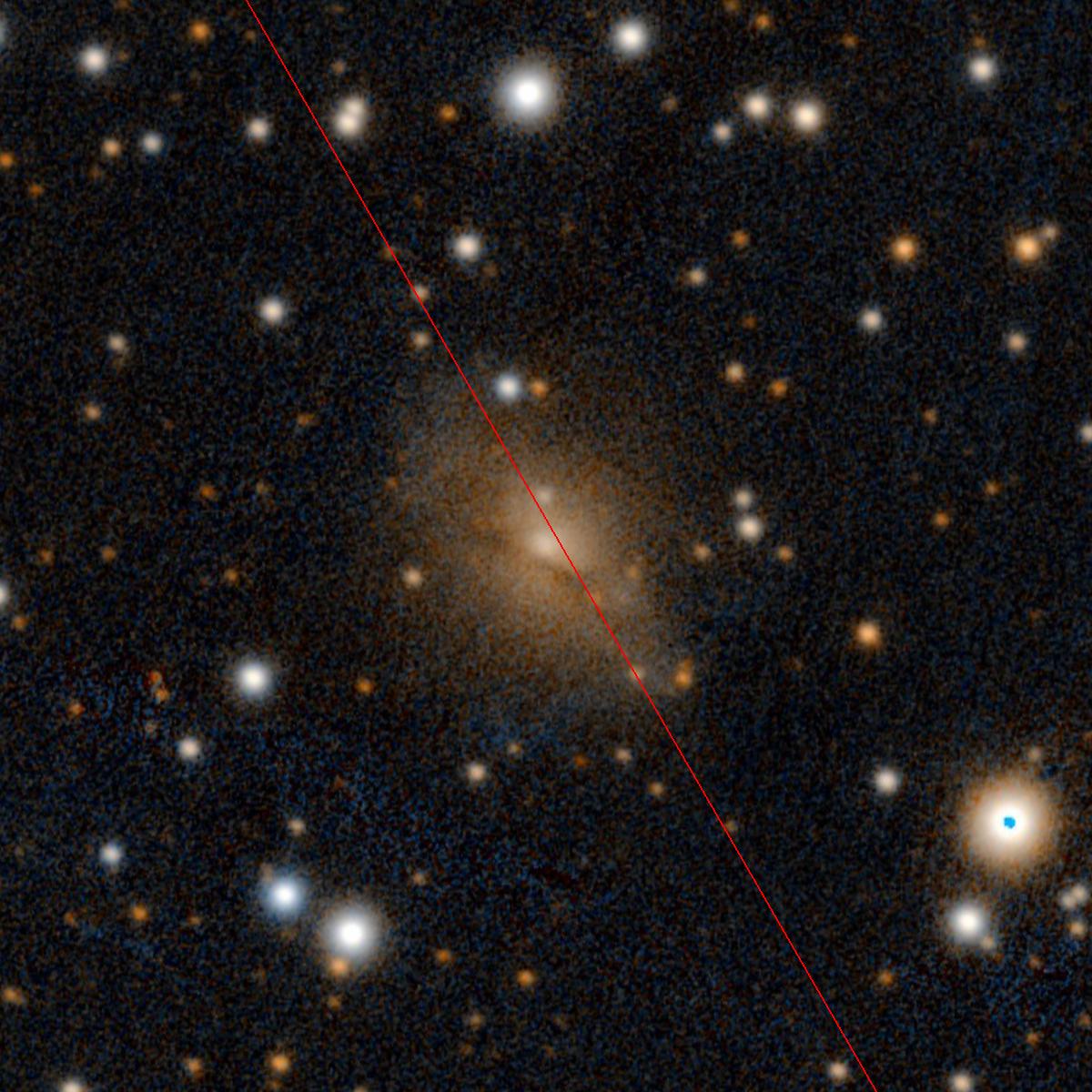}
\includegraphics[width=\linewidth]{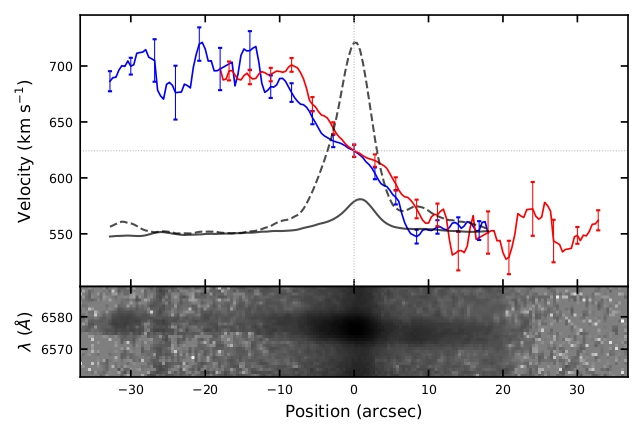}
\caption{PGC\,168301. 
\textit{Upper panel}: Long-slit position overlaid on the $2^{\prime}\times2^{\prime}$ Pan-STARRS1 image. 
\textit{Bottom panel}: Rotation curve derived from the long-slit spectra. 
The blue curve shows the barycentric velocity along the slit. The red line is the mirrored rotation curve with respect to the galaxy center. 
The relative continuum and emission-line intensities are indicated by the solid and dashed black lines, respectively. 
The extracted H$\alpha$ emission-line profile is shown below.}
\label{fig:pgc168301}
\end{figure}

The \HI{} source ID$_\mathrm{FASHI}=20230038548$ with heliocentric $cz_{\sun} = 627$~\kms{} is located $40^{\prime\prime}$ south of the PGC\,168301 with no known velocity. 
However, because of the low galactic latitude of $b=-03.039^\circ$, it appears to be a fairly distant object. 
To confirm this identification, we performed long-slit observations at the 6-m telescope of the Special Astrophysical Observatory of the Russian Academy of Sciences with the multi-mode focal reducer SCORPIO-2~\citep{2011BaltA..20..363A}. 
The VPHG1200@540 grating provides spectral coverage in the range 3650--7300\,\AA{} and with a slit of $1^{\prime\prime}$ gives a spectral resolution of 5.2\,\AA{}.
In November 2025, an exposure of 900~s in total was obtained at a position angle of $150.4^\circ$, as shown in the upper panel of Fig.~\ref{fig:pgc168301}.

We performed the data reduction using standard procedures implemented in an IDL-based pipeline, as described in detail by \citet{2018MNRAS.478.3386E}.
To improve the signal-to-noise ratio, we summed the signal within a moving window of $2\farcs4$. 
Line-of-sight velocities along the slit were measured by fitting Gaussian profile to the H$\alpha$ emission line. 
The barycentric correction\footnote{\url{https://docs.astropy.org/en/stable/coordinates/velocities.html}} was applied using the Astropy package~\citep{2013A&A...558A..33A}. 
The bottom panel of Fig.~\ref{fig:pgc168301} presents the resulting data. 
We measured the systemic velocity of PGC\,168301 of $cz_{\sun} = 624.2 \pm 5.4$~\kms{} at the maximum of the H$\alpha$ emission, which is in excellent agreement with the FASHI velocity of 627~\kms{}.

\section{Discussion and conclusion}

\begin{figure}
\centering
\includegraphics[width=\linewidth]{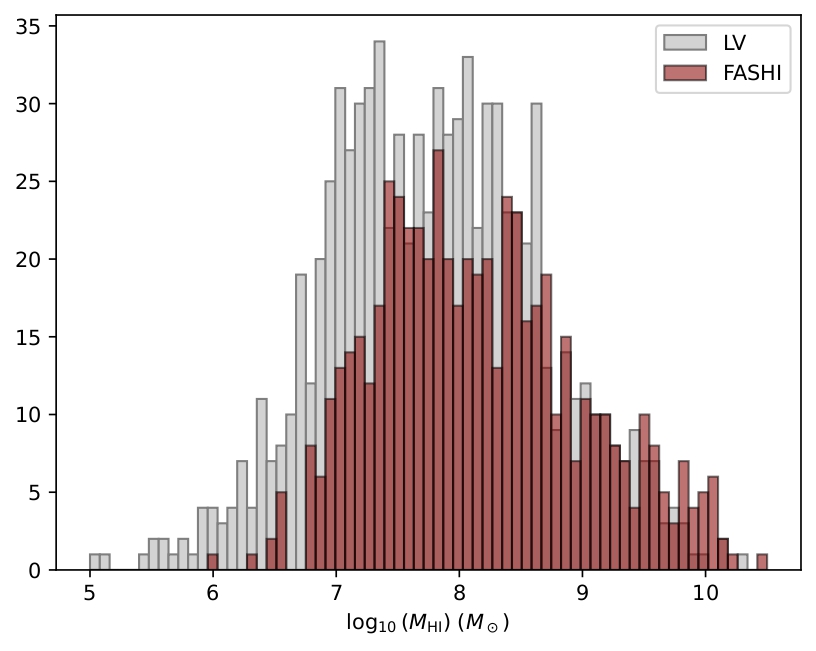}
\caption{
Histogram of the \HI{}-mass distribution in our sample of nearby FASHI sources compared to the \HI{} reservoir distribution in Local Volume galaxies.}
\label{fig:HIhist}
\end{figure}

We carried out visual identification of 662 \HI{}-sources from the FASHI survey~\citep{2024SCPMA..6719511Z} with velocities $V_\mathrm{LG} \leq 1150$~\kms{}, falling within the extended region up to about 16~Mpc of the Local Volume.
This corresponds to 636 unique objects, because some sources duplicate the same galaxies, usually the largest ones, such as M\,51 and M\,101.
In addition to the identifications listed in the FASHI catalog, we added 118 galaxies, including 71 new, previously uncatalogued objects.
We note that automatic identification with the SDSS photometric catalog without known redshifts and ignoring galaxy morphology is extremely unreliable in the nearby Universe. 
We rejected 48\% of such identifications. 
Otherwise, identifications with the ALFALFA~\citep{2005AJ....130.2598G} and SGA~\citep{2023ApJS..269....3M} catalogs are good. We made corrections in only three cases. 
For details and descriptions, see Section~\ref{sec:inspection}.

A comparison of the hydrogen mass distribution of the FASHI galaxies in our sample ($D\lesssim16$~Mpc) and the Local Volume galaxies ($D\lesssim12$ Mpc) is shown in Fig.~\ref{fig:HIhist}.
It can be seen that in the nearby Universe the FASHI survey is complete up to approximately $M_\mathrm{HI}\approx3$--$4\times10^7$~$M_\odot$.
The Local Volume allows us to advance up to $M_\mathrm{HI}=10^7$~$M_\odot$ due to specialized HI surveys of nearby dwarf galaxies~\citep{2025AJ....170...23N}.
Another option for increasing completeness is to search for weak signals in the \HI{}-maps from faint galaxies that were not detected by automatic detection algorithms~\citep{2025RAA....25l5012N}.

113 FASHI sources (105 unique objects) fall directly into the Local Volume ($V_\mathrm{LG} \leq 550$~\kms{}).
Among them, we discovered two new dwarf galaxies ID$_\mathrm{FASHI} = 20230063962$ and ID$_\mathrm{FASHI} = 20230001478$ in addition to the 20 previously found in FASHI survey by \citet{2024A&A...684L..24K}.
As noted by \citet{2024A&A...684L..24K}, there is significant room for discovery of new nearby galaxies, because the FASHI catalog is limited from below by a velocity of $V_\mathrm{h}=200$~\kms{}.

\begin{figure}
\centering
\includegraphics[width=\linewidth, bb=1 1 400 305, clip]{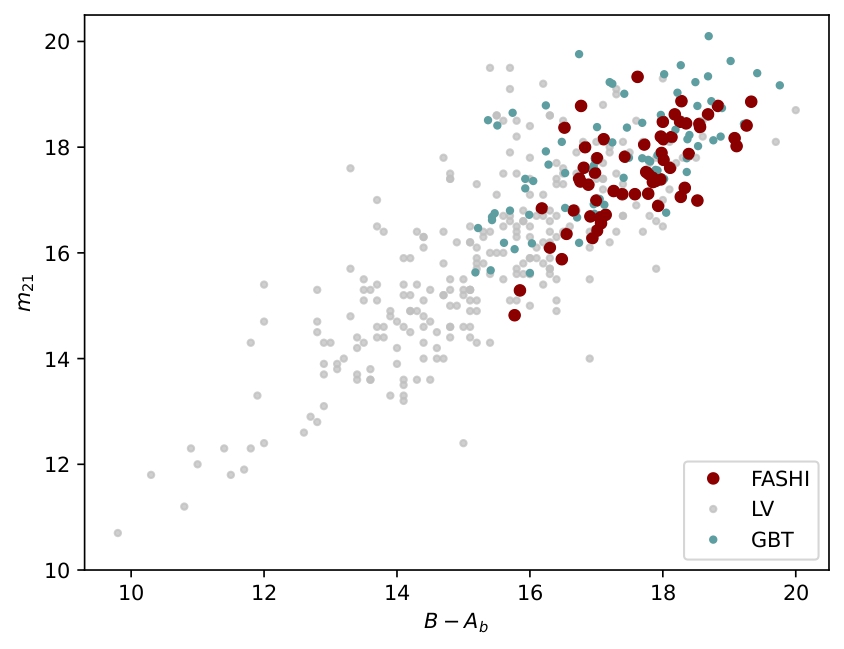}
\caption{Distribution of new discovered galaxies (dark red) according to $m_{21}$ magnitude and $B$-magnitude corrected for Galactic extinction combined with 266 late-type $T = [9, 10]$ dwarfs in the Local Volume with TRGB distance estimates (gray) and with 77 dwarf galaxies detected with the GBT (blue).}
\label{fig:B_m21}
\end{figure}

As already noted, during visual inspection, we discovered 71 new dwarf galaxies (Fig.~\ref{fig:mosaic}).
All of them demonstrate ongoing star formation and irregular morphology.
They are typically faint galaxies with $g$-band magnitudes ranging from 16.4 to 19.2~mag, with a median value of 17.8~mag and median blue color $(g-r)=0.29$~mag.
The \HI{}-flux varies from 0.17 to 26.4~Jy\,\kms{} with a median of $S_\mathrm{HI}=718$~mJy\,\kms{}.
Figure~\ref{fig:B_m21} presents their distribution according to $m_{21}$ and $B$ magnitudes in comparison with 266 LV galaxies lying outside the zone of high extinction ($A_b > 2\fm0$).
The figure also shows the distribution of 77 dwarf galaxies observed in the \HI{}-survey of recently discovered nearby dwarf galaxies with the 100-meter Green Bank Telescope~\citep{2025AJ....170...23N}.
As can be seen, the discovered galaxies follow the general trend of nearby LV galaxies and, as expected, are concentrated near the detection limit.

\begin{figure}
\centering
\includegraphics[width=0.9\textwidth]{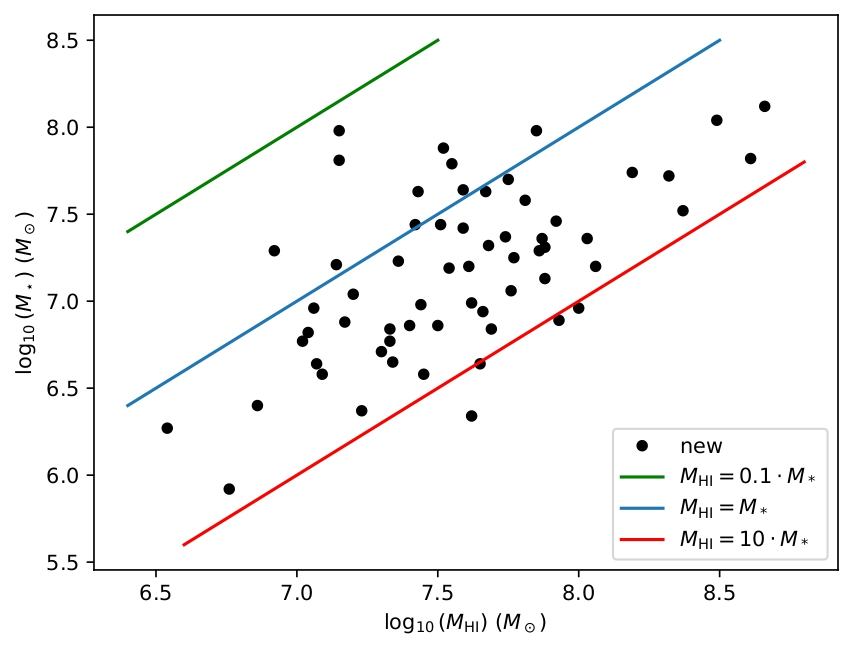}
\caption{Distribution of 61 new dwarf galaxies according to $M_\mathrm{HI}$ and $M_\star$ for which photometry and stellar mass estimate were performed. Three diagonal lines indicate the hydrogen-to-stellar ratio $\mu = M_\mathrm{HI}/M_\star$ equal to 0.1, 1, and 10.}
\label{fig:MHI_M*}
\end{figure}

The stellar masses of the galaxies discovered in this work span the range from $2.2\times10^6$ to $9.5\times10^7$~$M_\odot$, with a median value of $1.3\times10^7$~$M_\odot$. 
Their neutral hydrogen masses range from $8.7\times10^6$ to $9.5\times10^8$~$M_\odot$, with a median of $M_\mathrm{HI}=3.7\times10^7$~$M_\odot$.
As can be seen in Figure~\ref{fig:MHI_M*}, the vast majority of these systems are gas-dominated, with a median hydrogen-to-stellar mass ratio $M_\mathrm{HI}/M_\star = 3.4$.


\begin{figure}
\centering
\includegraphics[width=0.9\textwidth]{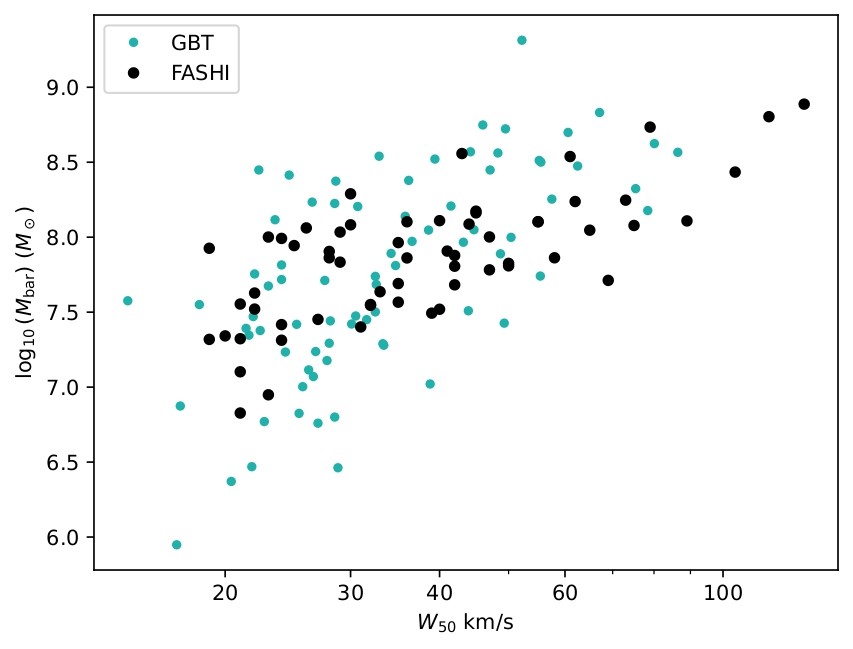}
\caption{Distribution of 61 new dwarf galaxies according to baryonic mass~$M_\mathrm{bar}$ and observed line width~$W_{50}$ for which photometry and stellar mass estimate were performed. Distribution of 77 dwarf galaxies detected with GBT is also shown.}
\label{fig:BTFR}
\end{figure}

As expected, low-mass systems dominate among the newly discovered galaxies. The internal kinematics (without corrections for inclination or turbulent gas motions) is characterized by a median $W_{50} = 37$~\kms{}, with a minimum $W_{50}$-value of 15~\kms{}. Interestingly, the sample also includes relatively massive systems with $W_{50}$ as high as $105$~\kms{}. Combining photometric and \HI{} data, we estimate the total baryonic mass as the sum of the stellar and gas components, $M_\mathrm{bar} = M_\star + M_\mathrm{gas}$, where the gas mass is taken as $M_\mathrm{gas} = 1.4 M_\mathrm{HI}$ to account for the contribution of helium and heavier elements. Figure~\ref{fig:BTFR} presents the simplified baryonic Tully–Fisher relation for the combined sample of nearby dwarf galaxies discovered in the FASHI survey and those observed with the GBT~\citep{2025AJ....170...23N}. It is evident that the two samples exhibit remarkably similar behavior, despite their fundamentally different selection strategies: the FASHI sample is based on a blind \HI{} survey, whereas the GBT sample comprises galaxies identified in deep optical imaging.


\begin{acknowledgements}
This work was supported by the Russian Science Foundation grant \textnumero~24--12--00277. 

This research has made use of ``Aladin sky atlas'' developed at CDS, Strasbourg Observatory, France.

We acknowledge the usage of the HyperLeda~\footnote{\url{https://leda.univ-lyon1.fr/}} database.

Observations with the SAO RAS telescopes are supported by the Ministry of Science and Higher Education of the Russian Federation. The renovation of telescope equipment is currently provided within the national project "Science and universities".
\end{acknowledgements}

\bibliographystyle{raa}
\bibliography{art}

\appendix

\newpage
\section{List of the discovered galaxies}
\label{app:ListNew}

The list of new galaxies is presented in Table~\ref{tab:list_new}.
In addition, it includes galaxies with the line-of-sight velocities determined for the first time.
For these galaxies we provide a common name.
The table also contains 20 objects (named FASHI0237+38, etc.) found previously by \citet{2024A&A...684L..24K}.
The columns are: 
(1) ID number of the radio source from the FASHI catalog; 
(2) the J2000.0 equatorial coordinates of the optical counterpart; 
(3) the common name for previously known galaxy, if exists; 
(4) radial velocity of the \HI{} source relative to the Local Group centroid (\kms{}); 
(5) the heliocentric velocity of \HI{} source (\kms{}); 
(6) the width of the \HI{} line at half intensity from the maximum (\kms{}); 
(7) the integrated \HI{} line flux (mJy\,\kms{}); 
(8) the distance (Mpc) to \HI{} source determined by its radial velocity using the NAM Calculator~\citep{2017ApJ...850..207S, 2020AJ....159...67K}; 
(9) the hydrogen mass (in solar masses) expressed as $\log(M_\mathrm{HI}/M_{\sun}) = 5.37 + \log(D_\mathrm{Mpc}) + \log(S_\mathrm{HI})$;
(10) the stellar mass (in solar masses) estimated using the $M/L$ relation justified by \citet{herr2016}.

{
\begin{footnotesize}
\setlength{\tabcolsep}{2.5pt}
\renewcommand{\arraystretch}{1.0}
\begin{ThreePartTable}

\begin{TableNotes}
\item[$(a)$] IR (unWISE)
\item[$(b)$] J124826.73+473905.5, J124834.98+474014.9, J124835.35+474043.5
\item[$(c)$] J225241.34+365719.8, J225253.47+365254.8
\end{TableNotes}

\begin{longtable}{c c l r r c r r c c}
\caption{List of new Local Volume candidate galaxies, also including galaxies from~\citet{2024A&A...684L..24K}.}\label{tab:list_new}\\
\hline\hline
FASHI ID &
J2000$_\mathrm{opt}$ &
\multicolumn{1}{c}{Name}	&
\multicolumn{1}{c}{$V_\mathrm{LG}$} &
\multicolumn{1}{c}{$cz$} &
\multicolumn{1}{c}{$W_{50}$} &
\multicolumn{1}{c}{$S_\mathrm{HI}$} &
\multicolumn{1}{c}{$D$} &
$\log M_\mathrm{HI}$ &
$\log M_\star$ \\

\cline{4-6}
 &
 &
 &
\multicolumn{3}{c}{km\,s$^{-1}$} &
\multicolumn{1}{c}{mJy\,km\,s$^{-1}$} &
\multicolumn{1}{c}{Mpc} &
$M_\odot$ &
$M_\odot$ \\

\hline
\endfirsthead

\caption{continue}\\
\hline
FASHI ID &
J2000$_{opt}$ &
\multicolumn{1}{c}{Name}	&
\multicolumn{1}{c}{$V_\mathrm{LG}$} &
\multicolumn{1}{c}{$cz$} &
\multicolumn{1}{c}{$W_{50}$} &
\multicolumn{1}{c}{$S_\mathrm{HI}$} &
\multicolumn{1}{c}{$D$} &
$\log M_\mathrm{HI}$ &
$\log M_\star$ \\

\cline{4-6}
 &
 &
 &
\multicolumn{3}{c}{km\,s$^{-1}$} &
\multicolumn{1}{c}{mJy km s$^{-1}$} &
\multicolumn{1}{c}{Mpc} &
$M_\odot$ &
$M_\odot$ \\
\hline
\endhead

\hline
\endfoot

\hline\hline
\insertTableNotes
\endlastfoot


20230057537	&	J000750.81+355757.3	  &	AGC000063, UGC00063   	&	713	&	438	&	39	&	3278.64	&	9.17	&	7.81  &  	  \\
20230062953	&	J011224.74+502338.2	  &		                    &	889	&	627	&	33	&	735.72	&	12.79	&	7.45  &    	  \\
20230030126	&	J012154.26+481408.1	  &		                    &  1128	&	873	&	29	&	539.13	&	15.55	&	7.49  &    	  \\
20230038719	&	J015517.09+573750.2	  &	PGC168257           	&	899	&	651	&	61	&  15412.10	&	13.23	&	8.80  &  	  \\
20230063123	&	J020120.24+512205.4	  &		                    &  1105	&	867	&	62	&	2543.40	&	15.61	&	8.16  &  	  \\
20230018432	&	J023053.16+391004.7	  &		                    &	821	&	624	&	43	&	365.34	&	12.49	&	7.13  &  	  \\
20230018411	&	J023115.48+390912.2	  &		                    &	701	&	504	&	23	&	465.11	&	11.15	&	7.13  &  	  \\
20230018469	&	J023423.90+391210.1	  &		                    &  1148	&	953	&	64	&	600.77	&	15.91	&	7.55  &  	  \\
20230057540	&	J023719.41+385600.1	  &	FASHI0237+38	        &	612	&	420	&	34	&	833.68	&	9.76	&	7.27  &  	  \\
20230003591	&	J024104.41$-$020640.4 &		                    &  1140	&  1093	&	36	&	1037.78	&	13.75	&	7.66  &  6.94 \\
20230004246	&	J024332.47$-$010837.6 &	PGC010311	            &  1014	&	965	&	30	&	1875.25	&	12.84	&	7.86  &  7.29 \\
20230016678	&	J024818.32+380054.0	  &		                    &	769	&	587	&	24	&	441.86	&	12.12	&	7.18  &  	  \\ 
20230021207	&	J025210.63+404702.3	  &	FASHI0252+40	        &	562	&	375	&	28	&	499.55	&	9.18	&	7.00  &  	  \\
20230018579	&	J025457.17+391524.7	  &	PGC010993           	&	660	&	480	&	52	&	4265.59	&	10.89	&	8.08  &  	  \\
20230025996	&	J025644.24+441006.6	  &		                    &	965	&	773	&	25	&	639.75	&	14.23	&	7.48  &  	  \\
20230038096	&	J030125.60+563752.4	  &		                    &	782	&	567	&	104	&  26408.50	&	12.35	&	8.98  &  	  \\
20230025382	&	J030218.46+433945.5   &	FASHI0302+43         	&	476	&	289	&	21	&	409.60	&	8.15	&	6.81  &  	  \\
20230003596	&	J033143.14$-$020617.0 &	PGC135676           	&	826	&	826	&	28	&	990.61	&	11.81	&	7.51  &  7.44 \\
20230057482	&	J035054.07$-$021943.3 &		                    &	948	&	966	&	22	&	368.40	&	13.48	&	7.20  &  7.04 \\
20230003718	&	J040433.19$-$015655.3 &	PGC135708           	&	802	&	831	&	24	&	1106.37	&	12.21	&	7.59  &  7.64 \\
20230063962	&   J040551.55+623321.0   &                         &   463 &   266 &   20  &   934.63  &   8.60    &   7.21  &       \\
20230002481	&	J041810.56$-$032655.7 &		                    &	766	&	814	&	39	&	2439.39	&	12.10	&	7.92  &  	  \\
20230007241	&	J050215.16+311955.2	  &		                    &	778	&	720	&	96	&	1804.91	&	14.03	&	7.92  &  	  \\
20230007694 &   J050603.71+314019.4\tnote{$a$} &	            &   777	&	720 &	105	&  11731.90	&	14.12	&	8.74  &  	  \\
20230038924	&	J052635.79+575151.8	  &	PGC2571898           	&	931	&	784	&	179	&  67062.60	&	15.72	&	9.59  &  	  \\
20230032296	&	J053940.79+504611.7	  &		                    &  1041	&	927	&	41	&	678.10	&	17.74	&	7.70  &  	  \\
20230001173	&	J055036.78$-$045914.1 &		                    &	895	&  1029	&	32	&	946.08	&	16.44	&	7.78  &  	  \\
20230040865	&	J055333.70+605649.5	  &		                    &	761	&	614	&	35	&	641.74	&	13.91	&	7.46  &  	  \\
20230060134	&	J055510.77$-$053159.7 &		                    &	883	&  1022	&	35	&	956.52	&	16.37	&	7.78  &  	  \\
20230001478	&	J060407.71$-$043824.7 &		                    &	532	&	675	&	48	&	1476.82	&	12.39	&	7.73  &  	  \\
20230057490	&	J061001.06$-$053315.0 &		                    &	568	&	719	&	39	&	510.60	&	12.93	&	7.30  &  	  \\
20230001267	&	J062144.45$-$045023.2 &		                    &	599	&	756	&	40	&	876.55	&	13.48	&	7.57  &  	  \\
20230036147	&	J064150.60+542731.9	  &		                    &  1073	&	972	&	69	&	395.69	&	18.38	&	7.50  &  6.86 \\
20230032587	&	J070339.17+510557.8	  &	PGC2388360	            &  1148	&  1070	&	79	&   3305.02	&	19.90	&	8.49  &  8.04 \\
20230028686	&	J070538.10+463708.8	  &		                    &  1005	&	949	&	38	&	380.75	&	18.70	&	7.50  &  	  \\
20230029424	&	J071200.70+472449.3	  &	PGC2297731	            &  1000	&	943	&	44	&   408.75	&	18.61	&	7.52  &  7.88 \\
20230014509	&	J073846.34+364230.6	  &		                    &	817	&	824	&	29	&	676.93	&	17.13	&	7.67  &  7.63 \\
20230057605	&	J074553.26+565951.6	  &		                    &	786	&	692	&	34	&	1113.84	&	15.78	&	7.81  &  	  \\
20230037169	&	J074656.15+552625.7	  &		                    &  1070	&	984	&	45	&	972.77	&	19.16	&	7.92  &  7.46 \\
20230036878	&	J080521.43+550856.8	  &		                    &  1041	&	961	&	42	&	565.27	&	19.17	&	7.69  &  6.84 \\
20230041248	&	J090233.84+621158.1	  &	SDSSJ090233.95+621157.2	&	902	&	794	&	42	&	1009.56	&	17.04	&	7.84  &  7.19 \\
20230034376	&	J090654.84+525351.4	  &		                    &	703	&	644	&	65	&	700.95	&	14.67	&	7.55  &  7.79 \\
20230057566	&	J092633.85+481331.1	  &		                    &	695	&	663	&	26	&	280.15	&	14.58	&	7.15  &  7.98 \\
20230037149	&	J094301.11+552517.6	  &		                    &	882	&	811	&	21	&	324.77	&	16.75	&	7.33  &  6.77 \\
20230024758	&	J095006.22+430731.4	  &		                    &	610	&	606	&	25	&	1235.88	&	12.81	&	7.68  &  7.32 \\
20230028487	&	J095905.90+462652.4	  &		                    &	638	&	615	&	42	&	717.50	&	12.84	&	7.44  &  6.98 \\
20230035438	&	J100507.01+534708.0	  & SDSSJ100507.13+534708.2 &	784	&	722	&	73	&	2158.60	&	14.96	&	8.06  &  7.20 \\
20230024875	&	J102429.56+431359.2	  &	PGC2215970              &	574	&	567	&	75	&	2469.08	&	11.42	&	7.88  &  7.13 \\
20230014801	&	J102823.95+365102.7	  &		                    &	956	&	983	&	45	&	1173.86	&	19.05	&	8.00  &  6.96 \\
20230038358	&	J104210.74+570405.1	  &		                    &	857	&	773	&	36	&	1311.23	&	15.49	&	7.87  &  7.36 \\
20230057638	&	J104442.34+630557.1	  &		                    &  1079	&	964	&	25	&	6439.20	&	17.64	&	8.67  &  	  \\
20230040619	&	J104724.82+602344.9	  &	PGC2601584          	&  1127	&  1025	&	30	&	823.21	&	19.22	&	7.85  &  7.98 \\
20230057607	&	J104826.79+581502.8	  &		                    &  1099	&  1008	&	24	&	217.89	&	18.87	&	7.26  &  	  \\
20230032161	&	J104908.79+503421.0	  &	SDSSJ104908.62+503425.7	&	917	&	866	&	23	&	888.76	&	16.82	&	7.77  &  7.25 \\
20230017803	&	J110435.60+384414.0	  &	FASHI1104+38	        &	601	&	610	&	24	&	478.89	&	9.64	&	7.02  &  6.77 \\
20230004494	&	J111438.79$-$005114.7 &	PGC1133017          	&	773	&	973	&	89	&	1930.75	&	12.00	&	7.81  &  7.58 \\
20230001298	&	J112221.34$-$044640.0 &	PGC1053625	            &	784	&	995	&	116	&  10749.70	&	12.71	&	8.61  &  7.82 \\
20230057498	&	J113145.36$-$050733.4 &		                    &	839	&  1049	&	26	&	377.24	&	13.78	&	7.23  &       \\
20230032964	&	J120536.43+513257.1	  &		                    &  1089	&  1014	&	29	&	212.41	&	20.62	&	7.33  &       \\
20230027281	&	J120645.15+452938.1	  &	FASHI1206+45	        &	656	&	610	&	24	&	274.37	&	9.69	&	6.78  &       \\
20230142538	&	J120732.55+391839.5	  &		                    &  1073	&  1058	&	19	&	168.95	&	18.93	&	7.15  &  7.81 \\
20230032432	&	J120804.70+505355.7	  &		                    &  1098	&  1025	&	41	&	381.67	&	20.95	&	7.59  &  7.42 \\
20230029574	&	J121015.84+473426.4	  &	FASHI1210+47         	&	466	&	408	&	23	&	493.44	&	7.06	&	6.76  &  5.92 \\
20230027590	&	J121046.62+454323.6	  &	PGC2832109          	&  1004	&	955	&	28	&	325.16	&	18.77	&	7.43  &  7.63 \\
20230025550	&	J121933.42+434712.3	  &		                    &  1007	&	964	&	22	&	357.35	&	18.22	&	7.45  &       \\
20230021454	&	J122010.44+405242.6	  &	FASHI1220+40	        &	548	&	520	&	21	&	280.31	&	7.26	&	6.54  &  6.27 \\
20230140919	&	J122455.33+605918.0	  &	PGC040431               &	837	&	711	&	61	&	4062.98	&	14.78	&	8.32  &  7.72 \\
20230032541	&	J122731.15+510132.2	  &		                    &  1134	&  1053	&	29	&	384.06	&	21.42	&	7.62  &  6.99 \\
20230058904	&	J123315.24+354403.8	  &	FASHI1233+35, AGC226688	&	826	&	817	&	35	&	907.07	&	10.32	&	7.36  &  7.23 \\
20230021429	&	J123830.39+405149.2	  &	FASHI1238+40         	&	681	&	644	&	22	&	313.66	&	9.21	&	6.80  &       \\
20230011719	&	J124458.69+344103.1	  &		                    &	889	&	878	&	31	&	259.36	&	13.79	&	7.06  &  6.96 \\
20230057484	&	J124713.62$-$050703.1 &	PGC1049083	            &	874	&  1041	&	39	&	574.88	&	7.81	&	6.92  &  7.29 \\
20230029664 &   triplet\tnote{$b$}    &                         &   966 &   892 &   28  &   267.07  &   17.46   &   7.28  &       \\ 
20230038542	&	J125104.08+572244.4	  &	FASHI1251+57	        &	544	&	426	&	21	&	639.48	&	9.03	&	7.09  &  6.58 \\
20230058517	&	J125139.39+321049.9	  &	FASHI1251+32, AGC229439	&	851	&	849	&	19	&	483.01	&	10.15	&	7.07  &  6.64 \\
20230034346	&	J125411.19+525211.5	  &		                    &	761	&	661	&	27	&	368.51	&	13.01	&	7.17  &  6.88 \\
20230012237	&	J125752.61+352223.9	  &	SDSSJ125752.59+352224.1	&	992	&	971	&	28	&	178.10	&	14.36	&	6.94  &       \\
20230003668	&	J130059.63$-$020107.9 &	PGC135815            	&	986	&  1131	&	35	&	1058.81	&	15.16	&	7.76  &  7.06 \\
20230002525	&	J131228.84$-$032328.8 &		                    &	965	&  1108	&	62	&	1963.81	&	15.19	&	8.03  &  7.36 \\
20230007323	&	J131318.92+312449.0	  &	FASHI1313+31, AGC239104	&	812	&	801	&	20	&	500.94	&	9.69	&	7.04  &  6.82 \\
20230061057 &   J132212.50+323302.4   & [SMB88]\,2307           &   867	&	844	&	29	&	347.50	&	10.54	&	6.96  &       \\
20230061148	&	J132834.61+330849.1	  &	FASHI1328+33, AGC239114	&	784	&	754	&	33	&	1295.47	&	9.60	&	7.45  &  6.58 \\
20230008631	&	J133009.14+321716.3	  &	FASHI1330+32, AGC239115	&	771	&	744	&	24	&	827.89	&	9.38	&	7.23  &  6.37 \\
20230021427	&	J133135.83+405215.2	  &		                    &  1096	&  1030	&	43	&	2800.84	&	18.85	&	8.37  &  7.52 \\
20230027451	&	J133212.50+453717.8	  &		                    &	758	&	671	&	47	&	1452.88	&	11.09	&	7.62  &  6.34 \\
20230057538	&	J133409.51+412754.3	  &		                    &  1139	&  1069	&	15	&	1216.46	&	20.54	&	8.08  &       \\
20230064037	&	J133418.55+625738.8	  &	FASHI1334+62        	&	549	&	392	&	35	&	924.12	&	9.92	&	7.33  &  6.84 \\
20230036458	&	J133555.68+544431.2	  &	FASHI1335+54        	&	485	&	359	&	21	&	549.40	&	7.46	&	6.86  &  6.40 \\
20230018394	&	J133945.15+390807.7	  &	FASHI1339+39        	&	745	&	682	&	25	&	1028.65	&	9.96	&	7.38  &       \\
20230063286	&	J135407.20+534713.2	  &	FASHI1354+53        	&	428	&	297	&	22	&	335.66	&	6.43	&	6.51  &       \\
20230057555	&	J141608.78+471408.9	  &		                    &	768	&	650	&	32	&	650.15	&	11.90	&	7.34  &  6.65 \\
20230024668	&	J142012.21+430110.9	  &	FASHI1420+43         	&	733	&	628	&	32	&	502.65	&	10.86	&	7.14  &  7.21 \\
20230017986	&	J144506.71+385327.7	  &		                    &	780	&	674	&	58	&	1452.41	&	10.93	&	7.61  &  7.20 \\
20230034779	&	J144509.94+531259.0	  &		                    &	994	&	838	&	22	&	405.37	&	16.19	&	7.40  &  6.86 \\
20230062132	&	J144604.49+413751.6	  &		                    &	761	&	645	&	50	&	1605.30	&	10.91	&	7.65  &  6.64 \\
20230035557	&	J145844.06+535247.6	  &		                    &  1050	&	885	&	50	&	3933.77	&	17.03	&	8.43  &  	  \\
20230057620	&	J152242.07+623408.4	  &		                    &	694	&	495	&	47	&	1737.85	&	11.58	&	7.74  &  7.37 \\
20230057621 &   J152857.85+615206.1   & PGC055203               &   720 &   520 &   40  &   1682.61 &   11.91   &   7.75  &  7.70 \\
20230030720	&	J152937.87+485103.6	  &	PGC3097758           	&	642	&	473	&	50	&	1049.76	&	10.34	&	7.42  &  7.44 \\
20230063225	&	J153225.66+522438.2	  &		                    &	920	&	741	&	24	&	280.54	&	15.02	&	7.17  &  	  \\
20230057618	&	J154852.72+620838.4	  &	PGC3086677          	&  1060	&	851	&	130	&	6957.91	&	16.78	&	8.66  &  8.12 \\
20230057619	&	J155532.59+623618.0	  &		                    &  1038	&	825	&	55	&	1367.43	&	16.22	&	7.93  &  6.89 \\
20230035546	&	J160213.76+535200.4	  &		                    &	973	&	773	&	55	&	1304.04	&	15.65	&	7.88  &  7.31 \\
20230034243	&	J162444.47+524723.3	  &		                    &  1147	&	937	&	40	&	260.40	&	18.06	&	7.30  &  6.71 \\
20230016588	&	J165328.30+375652.8	  &	PGC2113799          	&  1050	&	854	&	104	&	2606.28	&	15.85	&	8.19  &  7.74 \\
20230062028	&	J222516.63+400350.7	  &		                    &  1034	&	724	&	47	&	610.47	&	12.68	&	7.36  &  	  \\
20230062297	&	J223930.21+430659.8	  &		                    &	725	&	415	&	37	&	718.08	&	9.19	&	7.15  &  	  \\
20230014875 &	doublet\tnote{$c$}	  &		                    &  1069 &	767	&	29	&	2914.41	&	12.73	&	8.05  &  	  \\
20230027330	&	J231207.96+453121.8	  &		                    &  1116	&	812	&	36	&	741.76	&	14.55	&	7.57  &  	  \\   
\end{longtable}
\end{ThreePartTable}
\end{footnotesize}
}

\newpage
\section{Photometry of new galaxies}
\label{app:Photometry}

The results of surface brightness photometry For 62 of the new LV candidate galaxies are presented in Table~\ref{tab:photometry}. The columns of the table contain: 
(1) ID number of the \HI{} source in the FASHI catalog; 
(2) the asymptotic total magnitude $m$ in the $g$ band; 
(3)-(4) $r_{50}$ and $r_{80}$ radius in the $g$ band
(5) the asymptotic total magnitude $m$ in the $r$ band; 
(6)-(7) $r_{50}$ and $r_{80}$ radius in the $r$ band
(8)-(9) $B$ and $V$ magnitudes; 
(10) the $m_{21}$ magnitude.

{
\begin{footnotesize}
\setlength{\tabcolsep}{3pt}
\renewcommand{\arraystretch}{1.0}

\begin{ThreePartTable}

\begin{TableNotes}
\item[ $(a)$ ] Values taken from the Siena Galaxy Atlas~\citep[SGA,][]{2023ApJS..269....3M}.
\end{TableNotes}

\begin{longtable}{l c r r c r r c c c} 
\caption{Photometric values for some new Local Volume candidate galaxies.} \label{tab:photometry}\\
\hline\hline
\multicolumn{1}{c}{FASHI ID} &
$m_\mathrm{g,tot}$ &
$r_{g,50}$ &
$r_{g,80}$ &
$m_\mathrm{r,tot}$ &
$r_{r,50}$ &
$r_{r,80}$ &
$B$ &
$V$ &
$m_{21}$ \\

 &
mag &
\multicolumn{1}{c}{$''$} &
\multicolumn{1}{c}{$''$} &
mag &
\multicolumn{1}{c}{$''$} &
\multicolumn{1}{c}{$''$} &
mag &
mag &
mag \\
\hline
\endfirsthead

\caption{continue}\\
\hline
\multicolumn{1}{c}{FASHI ID} &
$m_\mathrm{g,tot}$ &
$r_{g,50}$ &
$r_{g,80}$ &
$m_\mathrm{r,tot}$ &
$r_{r,50}$ &
$r_{r,80}$ &
$B$ &
$V$ &
$m_{21}$ \\

 &
mag &
\multicolumn{1}{c}{$''$} &
\multicolumn{1}{c}{$''$} &
mag &
\multicolumn{1}{c}{$''$} &
\multicolumn{1}{c}{$''$} &
mag &
mag &
mag \\
\hline
\endhead

\hline
\endfoot

\hline\hline
\insertTableNotes
\endlastfoot

20230003591             & 17.73 &  6.7 &  9.8 & 17.49 &  6.9 & 10.2 & 18.00 & 17.60 & 17.36 \\
20230004246             & 16.95 & 15.7 & 26.9 & 16.63 & 16.2 & 27.9 & 17.26 & 16.78 & 16.72 \\
20230003596             & 16.59 & 12.1 & 31.9 & 16.21 & 13.3 & 34.9 & 16.94 & 16.39 & 17.41 \\
20230057482             & 18.18 &  7.3 &  9.8 & 17.72 &  7.8 & 10.3 & 18.57 & 17.94 & 18.48 \\
20230003718             & 17.11 & 18.1 & 29.7 & 16.46 & 18.8 & 31.0 & 17.60 & 16.77 & 17.29 \\
20230036147             & 19.10 &  3.6 &  5.5 & 18.69 &  3.9 &  6.2 & 19.49 & 18.91 & 18.41 \\
20230032587\tnote{$a$}  & 16.17 &  7.4 &      & 15.81 &  7.5 &      & 16.54 & 16.01 & 16.10 \\
20230029424\tnote{$a$}  & 16.49 &  6.4 &      & 16.11 &  5.4 &      & 16.86 & 16.31 & 18.37 \\
20230014509             & 17.25 & 10.0 & 16.2 & 16.77 & 10.8 & 20.6 & 17.68 & 17.02 & 17.82 \\
20230057605             & 17.84 &  9.3 & 14.3 &       &      &      &       &       &       \\
20230037169             & 17.66 &  6.1 & 10.1 & 17.26 &  6.2 & 10.5 & 18.05 & 17.47 & 17.43 \\
20230036878             & 18.95 &  4.6 &  7.7 & 18.63 &  4.5 &  7.4 & 19.29 & 18.80 & 18.02 \\
20230041248             & 17.84 &  3.9 &  7.6 & 17.51 &  3.8 &  7.7 & 18.19 & 17.69 & 17.39 \\
20230034376             & 16.62 & 18.3 & 32.3 & 16.10 & 18.0 & 31.1 & 17.07 & 16.37 & 17.79 \\
20230057566             & 16.33 & 27.9 & 52.5 & 15.75 & 28.5 & 52.9 & 16.82 & 16.05 & 18.78 \\
20230037149             & 18.42 &  5.4 &  8.7 & 18.23 &  5.4 &  8.7 & 18.71 & 18.34 & 18.62 \\
20230024758             & 16.94 & 12.0 & 21.8 & 16.59 & 12.0 & 21.8 & 17.30 & 16.78 & 17.17 \\
20230028487             & 17.71 & 11.1 & 15.9 & 17.39 & 11.5 & 16.4 & 18.05 & 17.56 & 17.76 \\
20230035438             & 16.85 & 14.9 & 20.3 & 16.76 & 14.8 & 19.9 & 17.10 & 16.81 & 16.56 \\
20230024875\tnote{$a$}  & 16.77 &  8.8 &      & 16.54 &  8.8 &      & 17.06 & 16.67 & 16.42 \\
20230014801             & 18.11 & 11.0 & 15.8 & 17.97 & 10.7 & 15.6 & 18.38 & 18.05 & 17.23 \\
20230038358             & 17.07 & 12.8 & 20.0 & 16.78 & 12.9 & 19.9 & 17.40 & 16.94 & 17.11 \\
20230040619\tnote{$a$}  & 16.90 & 10.6 &      & 16.75 & 11.3 &      & 16.85 & 16.24 & 17.61 \\
20230032161             & 17.48 & 10.3 & 15.4 & 17.20 & 10.0 & 14.8 & 17.80 & 17.35 & 17.53 \\
20230017803             & 17.67 &  8.5 & 15.5 & 17.33 &  9.6 & 16.6 & 18.03 & 17.51 & 18.20 \\
20230004494\tnote{$a$}  & 16.66 & 13.2 &      & 16.16 & 13.9 &      & 17.07 & 16.39 & 16.69 \\
20230001298\tnote{$a$}  & 15.63 & 67.0 &      & 15.31 & 66.3 &      & 15.95 & 15.46 & 14.82 \\
20230142538             & 17.26 &  9.1 & 17.2 & 16.70 &  9.8 & 18.1 & 17.73 & 16.99 & 19.33 \\
20230032432             & 18.03 &  4.3 &  8.7 & 17.50 &  5.9 &  9.8 & 18.43 & 17.83 & 18.45 \\
20230029574             & 18.86 &  7.7 & 13.0 & 18.60 &  8.3 & 13.9 & 19.17 & 18.74 & 18.17 \\
20230027590             & 17.74 & 10.3 & 20.4 & 17.16 & 20.4 & 22.4 & 18.23 & 17.46 & 18.62 \\
20230021454             & 18.53 &  6.3 & 11.2 & 18.12 &  6.5 & 11.8 & 18.92 & 18.34 & 18.78 \\
20230140919             & 16.18 &  8.3 & 14.6 & 15.85 &  8.5 & 14.9 & 16.53 & 16.03 & 15.88 \\
20230032541             & 18.32 &  6.6 & 10.6 & 18.16 &  6.8 & 10.9 & 18.60 & 18.25 & 18.44 \\
20230058904\tnote{$a$}  & 16.67 &  6.8 &      & 16.33 &  7.6 &      & 17.02 & 16.51 & 17.51 \\
20230011719             & 17.98 &  7.4 & 10.2 & 17.64 &  7.2 & 10.7 & 18.34 & 17.82 & 18.87 \\
20230057484\tnote{$a$}  & 16.49 & 10.7 &      & 15.98 & 11.2 &      & 16.91 & 16.22 & 18.00 \\
20230038542             & 17.71 &  9.7 & 14.4 & 17.46 & 10.1 & 15.2 & 18.02 & 17.60 & 17.89 \\
20230058517             & 17.90 &  9.1 & 15.2 & 17.63 & 10.1 & 16.4 & 18.19 & 17.75 & 18.19 \\
20230034346             & 17.95 &  7.2 & 10.2 & 17.64 &  7.3 & 10.5 & 18.29 & 17.81 & 18.48 \\
20230003668             & 17.65 & 12.0 & 19.5 & 17.40 & 12.3 & 20.6 & 17.93 & 17.51 & 17.34 \\
20230002525             & 16.88 & 14.2 & 21.9 & 16.64 & 14.9 & 23.1 & 17.15 & 16.75 & 16.67 \\
20230007323             & 17.70 & 15.0 & 23.1 & 17.32 & 15.2 & 23.2 & 18.05 & 17.50 & 18.15 \\
20230061148             & 17.57 &  8.6 & 13.0 & 17.43 &  9.0 & 13.2 & 17.81 & 17.48 & 17.12 \\
20230008631             & 17.92 &  6.0 & 10.1 & 17.83 &  6.4 & 10.6 & 18.14 & 17.86 & 17.61 \\
20230021427             & 16.70 & 16.5 & 24.7 & 16.55 & 16.0 & 24.1 & 16.98 & 16.64 & 16.28 \\
20230027451             & 18.34 &  7.4 & 13.2 & 18.25 &  7.1 & 12.8 & 18.59 & 18.30 & 16.99 \\
20230064037             & 17.52 & 14.6 & 22.0 & 17.19 & 15.2 & 22.7 & 17.87 & 17.37 & 17.49 \\
20230036458             & 17.48 &  9.7 & 15.6 & 17.33 &  9.8 & 15.3 & 17.76 & 17.42 & 18.05 \\
20230057555             & 18.13 & 12.4 & 17.5 & 17.88 & 12.5 & 17.7 & 18.44 & 18.02 & 17.87 \\
20230024668             & 16.80 &  6.3 & 11.3 & 16.47 &  6.8 & 12.2 & 17.15 & 16.65 & 18.15 \\
20230017986             & 16.71 & 15.1 & 23.6 & 16.42 & 15.0 & 23.6 & 17.04 & 16.58 & 16.99 \\
20230034779             & 18.29 &  9.6 & 14.9 & 18.04 &  9.9 & 15.7 & 18.60 & 18.18 & 18.38 \\
20230062132             & 17.71 &  8.1 & 11.8 & 17.56 &  8.4 & 12.1 & 17.99 & 17.65 & 16.89 \\
20230057620             & 16.41 & 11.3 & 15.2 & 16.12 & 11.6 & 15.4 & 16.74 & 16.28 & 16.80 \\
20230057621\tnote{$a$}  & 15.88 &  8.4 &      & 15.51 &  9.4 &      & 16.25 & 15.71 & 16.84 \\
20230030720             & 16.42 &  7.4 & 15.0 & 16.00 &  8.3 & 18.8 & 16.82 & 16.22 & 17.35 \\
20230057618             & 15.54 &  6.6 & 12.0 & 15.19 &  6.8 & 12.9 & 15.90 & 15.38 & 15.29 \\
20230057619             & 18.03 & 10.1 & 15.6 & 17.85 &  9.8 & 15.0 & 18.32 & 17.95 & 17.06 \\
20230035546             & 17.28 &  9.3 & 17.4 & 16.97 & 10.0 & 19.0 & 17.62 & 17.14 & 17.11 \\
20230034243             & 19.10 &  6.4 &  8.8 & 18.79 &  6.6 &  9.4 & 19.44 & 18.96 & 18.86 \\   
20230016588             & 16.27 &  7.4 & 14.0 & 15.95 &  7.6 & 13.9 & 16.61 & 16.12 & 16.36 \\
\end{longtable}
\end{ThreePartTable}
\end{footnotesize}
}

\newpage
\section{List of the empty fields}
\label{app:Empty}

Our sample contains \HI{} sources that could not be reliably identified with any optical counterpart. These cases are presented in Table~\ref{tab:list_empty}.
The columns of the table contain: 
(1) ID number of the \HI{} source in the FASHI catalog; 
(2) the equatorial coordinates of the \HI{} source; 
(3) radial velocity of the \HI{} source relative to the Local Group centroid; 
(4) the heliocentric velocity of \HI{} source; 
(5) the \HI{} line-width at half intensity from the maximum; 
(6) the integrated \HI{} line flux; 
(7) the distance to \HI{} source determined by its radial velocity using the NAM Calculator~\citep{2017ApJ...850..207S, 2020AJ....159...67K}; 
(8) the hydrogen mass; 
(9) galactic latitude; 
(10) galactic extinction in $B$-band;
(11) galaxy group in whose virial zone the \HI{} source falls; 
(12) some notes on the \HI{} source in terms of possible candidates with corresponding coordinates, \citet{2024A&A...684L..24K} given names.

{
\begin{footnotesize}
\setlength{\tabcolsep}{1pt}
\renewcommand{\arraystretch}{1.0}

\begin{ThreePartTable}
\begin{TableNotes}
\item[ ] $p.c.$ marks \textit{possible candidate}
\end{TableNotes}
\begin{longtable}{c c c r c r r c c c c l}
\caption{List of empty fields.}\label{tab:list_empty}\\
\hline\hline
FASHI ID &
J2000 &
$V_\mathrm{LG}$ &
\multicolumn{1}{c}{$cz$} &
$W_{50}$ &
\multicolumn{1}{c}{$S_\mathrm{HI}$} &
\multicolumn{1}{c}{$D$} &
$\log M_\mathrm{HI}$ &
\multicolumn{1}{c}{$b$} &
\multicolumn{1}{c}{$A_B$} &
\multicolumn{1}{c}{Group} &
\multicolumn{1}{c}{Notes} \\

\cline{3-5}
 &
 &
\multicolumn{3}{c}{km s$^{-1}$} &
\multicolumn{1}{c}{mJy km s$^{-1}$} &
\multicolumn{1}{c}{Mpc} &
$M_\odot$ &
 &
 &
 &
 \\
\hline
\endfirsthead

\caption{continue}\\
\hline
FASHI ID &
J2000 &
$V_\mathrm{LG}$ &
\multicolumn{1}{c}{$cz$} &
$W_{50}$ &
\multicolumn{1}{c}{$S_\mathrm{HI}$} &
\multicolumn{1}{c}{$D$} &
$\log M_\mathrm{HI}$ &
\multicolumn{1}{c}{$b$} &
\multicolumn{1}{c}{$A_B$} &
\multicolumn{1}{c}{Group} &
\multicolumn{1}{c}{Notes} \\

\cline{3-5}
 &
 &
\multicolumn{3}{c}{km s$^{-1}$} &
\multicolumn{1}{c}{mJy km s$^{-1}$} &
\multicolumn{1}{c}{Mpc} &
$M_\odot$ &
 &
 &
 &
 \\
\hline
\endhead

\hline
\endfoot

\hline\hline
\insertTableNotes
\endlastfoot


20230063517	&	J015417.14+563255.4	  &	934	 &	686	&	22	&	329.04	&	13.65	&	7.16	&	-5.3	& 1.40 & 		    &   $p.c.$: J015418.95+563313.3 \\
20230039951	&	J020555.26+591949.3	  &	913	 &	668	&	31	&	2198.58	&	13.55	&	7.98	&	-2.2	& 3.08 & 		    &	$p.c.$ (ZTF): J020558.94+591919.8	\\
20230057471	&	J021201.69$-$013046.9 &	832	 &	757	&	122	&	524.53	&	10.26	&	7.11	&	-57.9	& 0.11 & 		    &		\\
20230063828	&	J022252.80+611110.9	  &	936	 &	697	&	105	&  21274.90 &	14.00	&	8.99	&	 0.2	& 3.64 & 		    &		\\
20230064273	&	J025604.50+645400.1	  &	948	 &	719	&	32	&	676.41	&	14.34	&	7.51	&	 5.2	& 4.53 & 		    &		\\
20230027599	&	J030019.48+454346.2	  & 1143 &	949	&	54	&	176.03	&	16.28	&	7.04	&	-11.5	& 1.03 & 		    &		\\
20230063775	&	J033053.68+604716.3	  &	 971 &	762	&	65	&	6097.33	&	14.97	&	8.51	&	 3.7	& 3.43 & 		    &	$p.c.$ (ZTF): J033055.27+604710.1	\\
20230036494	&	J040846.34+544620.5	  & 1130 &	955	&	68	&	5156.70	&	17.23	&	8.56	&	 2.2	& 8.46 & 		    &	$p.c.$ (unWISE): J040847.99+544616.3	\\
20230037287	&	J042503.86+553415.8	  & 1053 &	884	&	30	&	305.26	&	16.49	&	7.29	&	 4.4	& 3.97 & 		    &		\\
20230001605	&	J044037.39$-$042821.0 & 1130 & 1202	&	37	&	504.15	&	17.55	&	7.56	&	-31.0	& 0.23 & 		    &	wrong detection	\\
20230042234	&	J052509.04+325151.6	  &	669	 &	621	&	26	&	365.78	&	13.60	&	7.20	&	-1.6	& 2.35 & 		    &	$p.c.$: J052509.28+325214.6 \\ 
20230018960	&	J054031.82+392640.1	  &	684	 &	617	&	43	&	1240.39	&	14.06	&	7.76	&	 4.6	& 2.07 & 		    &	$p.c.$ (ZTF): J054031.59+392632.1 \\
20230057481	&	J054501.95$-$005827.9 &	793	 &	906	&	43	&	546.80	&	15.31	&	7.48	&	-15.2	& 2.31 & 		    &	$p.c.$ (unWISE): J054458.65$-$005742.3 	\\
20230025624	&	J064727.72+435152.8	  & 1140 & 1090	&	72	&	322.56	&	20.16	&	7.49	&	17.7	& 0.37 & 		    &	wrong detection \\
20230025619	&	J070400.47+435135.8	  & 1148 & 1105	&	119	&	293.80	&	20.38	&	7.46	&	20.6	& 0.44 & 	     	&	wrong detection	\\
20230022177	&	J070403.17+412440.6	  &	976	 &	946	&	31	&	232.04	&	18.58	&	7.27	&	19.8	& 0.32 & 		    &	wrong detection	\\
20230001598	&	J072309.47$-$042850.3 &	873	 & 1066	&	233	&	2610.39	&	17.53	&	8.27	&	 5.0	& 0.84 & 	        &		\\
20230011600	&	J080143.85+343249.5	  & 1056 & 1083	&	92	&	813.68	&	20.49	&	7.90	&	28.7	& 0.21 & 		    &	wrong detection	\\
20230004050	&	J080438.81$-$013146.0 &	914	 & 1115	&	65	&	644.64	&	18.49	&	7.71	&	15.5	& 0.12 & 		    &	$p.c.$ (unWISE): J080438.57$-$013314.1  \\
20230027144	&	J084333.16+452205.5	  & 1142 & 1121	&	90	&	731.72	&	21.45	&	7.90	&	38.2	& 0.09 &        	&	wrong detection	\\
20230010571	&	J085305.32+333538.3	  &	413	 &	458	&	33	&	252.91	&	11.11	&	6.87	&	38.9	& 0.11 &   NGC2683	&		\\
20230038509	&	J100333.66+571823.4	  & 1011 &	930	&	115	&	460.70	&	18.54	&	7.57	&	47.8	& 0.03 & 	     	&	wrong detection	\\
20230034885	&	J103845.85+532105.8	  & 1103 & 1040	&	35	&	4540.67	&	19.78	&	8.62	&	54.1	& 0.06 &   NGC3310	&	$p.c.$: J103850.39+531948.5 \& J103851.25+531934.9 \\
20230013075	&	J112033.04+355404.9	  &	808	 &	828	&	274	&	1192.66	&	14.41	&	7.76	&	68.8	& 0.07 & 		    &	wrong detection	\\
20230004629	&	J112719.42$-$004348.5 &	837	 & 1031	&	7	&	244.79	&	14.87	&	7.10	&	55.6	& 0.09 & 	    	&	wrong detection \\
20230004501	&	J114336.68$-$005031.7 &	929	 & 1116	&	73	&	722.97	&	14.68	&	7.56	&	57.6	& 0.07 & 		    &	wrong detection	\\
20230024885	&	J120501.30+431412.0	  &	957	 &	923	&	17	&	133.15	&	16.54	&	6.93	&	71.3	& 0.06 &   NGC4111	&		\\
20230024157	&	J120701.78+424200.0	  &	932	 &	900	&	65	&	2320.61	&	15.82	&	8.14	&	72.0	& 0.05 &   NGC4111	&		\\
20230024613	&	J120711.96+425855.5	  &	819	 &	786	&	165	&	6187.90	&	13.27	&	8.41	&	71.8	& 0.05 &   NGC4111	&	 	\\
20230024488	&	J120712.66+425440.6	  &	876	 &	843	&	215	&  11482.00	&	13.99	&	8.72	&	71.8	& 0.05 &   NGC4111  &		\\
20230024464	&	J120720.82+425340.0	  &	889	 &	856	&	112	&	7013.43	&	14.62	&	8.55	&	71.9	& 0.05 &   NGC4111	&		\\
20230025334	&	J120732.46+433748.3	  &	930	 &	893	&	130	&	163.57	&	15.93	&	6.99	&	71.3	& 0.05 &   NGC4111	&		\\
20230025721	&	J120920.62+435708.4	  &	964	 &	925	&	21	&	333.00	&	17.64	&	7.39	&	71.2	& 0.05 &   NGC4111  &		\\
20230024837	&	J120929.91+431151.5	  &	974	 &	939	&	69	&	1126.20	&	17.32	&	7.90	&	71.8	& 0.05 &   NGC4111  &	Close to UGC07146	\\ 
20230028812	&	J121917.32+464415.4	  &	460	 &	403	&	48	&	6695.34	&	6.89	&	7.87	&	69.4	& 0.06 &   NGC4258  &	FASHI1219+46a\\ 
20230028556	&	J121950.82+463011.7	  &	447	 &	391	&	35	&	3834.50	&	6.57	&	7.59	&	69.6	& 0.05 &   NGC4258	&   FASHI1219+46b \\ 
20230028794	&	J121952.48+463755.2	  &	445	 &	388	&	36	&	3657.25	&	6.54	&	7.57	&	69.5	& 0.06 &   NGC4258	&   FASHI1219+46c \\ 
20230021838	&	J123104.78+410828.3	  &	663	 &	629	&	36	&	6363.44	&	9.10	&	8.09	&	75.4	& 0.10 &   NGC4490	&   FASHI1231+4 \\ 
20230002137	&	J123511.40$-$034730.7 & 1003 & 1173	&	102	&	3079.05	&	15.41	&	8.23	&	58.8	& 0.12 &   NGC4546	&		\\
20230061084	&	J124137.33+325125.6	  &	607	 &	607	&	198	&	4589.68	&	7.10	&	7.74	&	83.9    & 0.07 &   NGC4631  &	A poss. HI trace of interaction in NGC4631 group \\
20230008373	&	J124154.31+320735.2	  &	667	 &	671	&	87	&  56330.10	&	7.69	&	8.89	&	84.6	& 0.06 &   NGC4631	&	Excess of blue star-like objects; XLSB dwarf	\\
20230058573	&	J124323.00+324321.0	  &	617  &	617	&	46	&	5939.58	&	7.21	&	7.86	&	84.1	& 0.05 &   NGC4631	&   FASHI1243+32, AGC229489 \\ 
20230022318	&	J125024.15+413110.4	  &	374	 &	328	&	31	&	355.38	&	4.39	&	6.21	&	75.6	& 0.04 &   NGC4736	&   FASHI1250+41 \\
20230022402	&	J125115.39+413540.0	  &	282	 &	235	&	43	&	664.84	&	3.17	&	6.20	&	75.5	& 0.04 &   NGC4736	&	FASHI1251+41; $p.c.$: dw1251+4138	\\ 
20230030525	&	J125708.36+483817.8	  &	894	 &	812	&	64	&	615.01	&	15.30	&	7.53	&	68.5	& 0.04 & 		    &	wrong detection	\\
20230114123	&	J131107.40+363101.7	  & 1094 & 1059	&	128	&	922.47	&	18.30	&	7.86	&	79.7	& 0.03 &   NGC5005	&	A poss. bridge between NGC5033 and NGC5002	\\
20230057534	&	J131133.86+362958.8	  & 1081 & 1047	&	259	&	3626.41	&	18.21	&	8.45	&	79.7	& 0.03 &   NGC5005	&	A poss. bridge between NGC5033 and NGC5002	\\
20230006251	&	J132247.22+301529.4	  &	 950 &	938	&	334	&	1243.63	&	14.18	&	7.77	&	82.4	& 0.05 & 	    	&		\\
20230021490	&	J133145.82+405357.9	  & 1094 & 1028	&	44	&	747.32	&	18.84	&	7.79	&	73.9	& 0.03 &   NGC5005	&   A poss. HI trace of interaction with 20230021427 \\ 

\end{longtable}
\end{ThreePartTable}
\end{footnotesize}
}

\newpage
\section{Spectra of the new galaxies}
\label{app:Spectra}

\HI{}-profiles of the new Local Volume candidate galaxies from FASHI are given in the same sequence as in Table~\ref{tab:list_new}. The image does not contain 20 objects (named FASHI0237+38, etc.) found previously by \citet{2024A&A...684L..24K}. 

\begin{figure}[h]
\centering
\includegraphics[width=\linewidth]{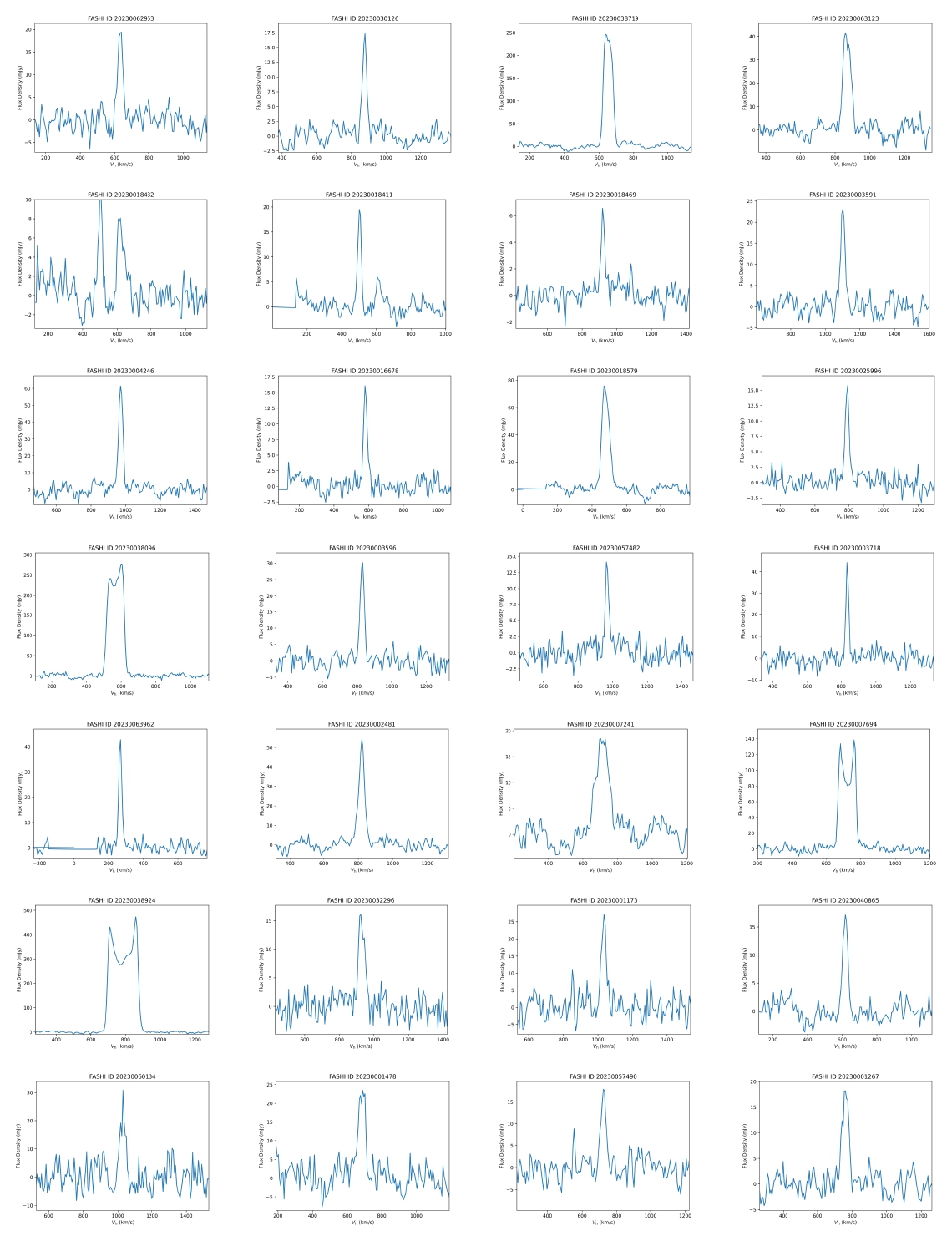}
\end{figure}

\addtocounter{figure}{-1}
\begin{figure}
\centering
\includegraphics[width=\linewidth]{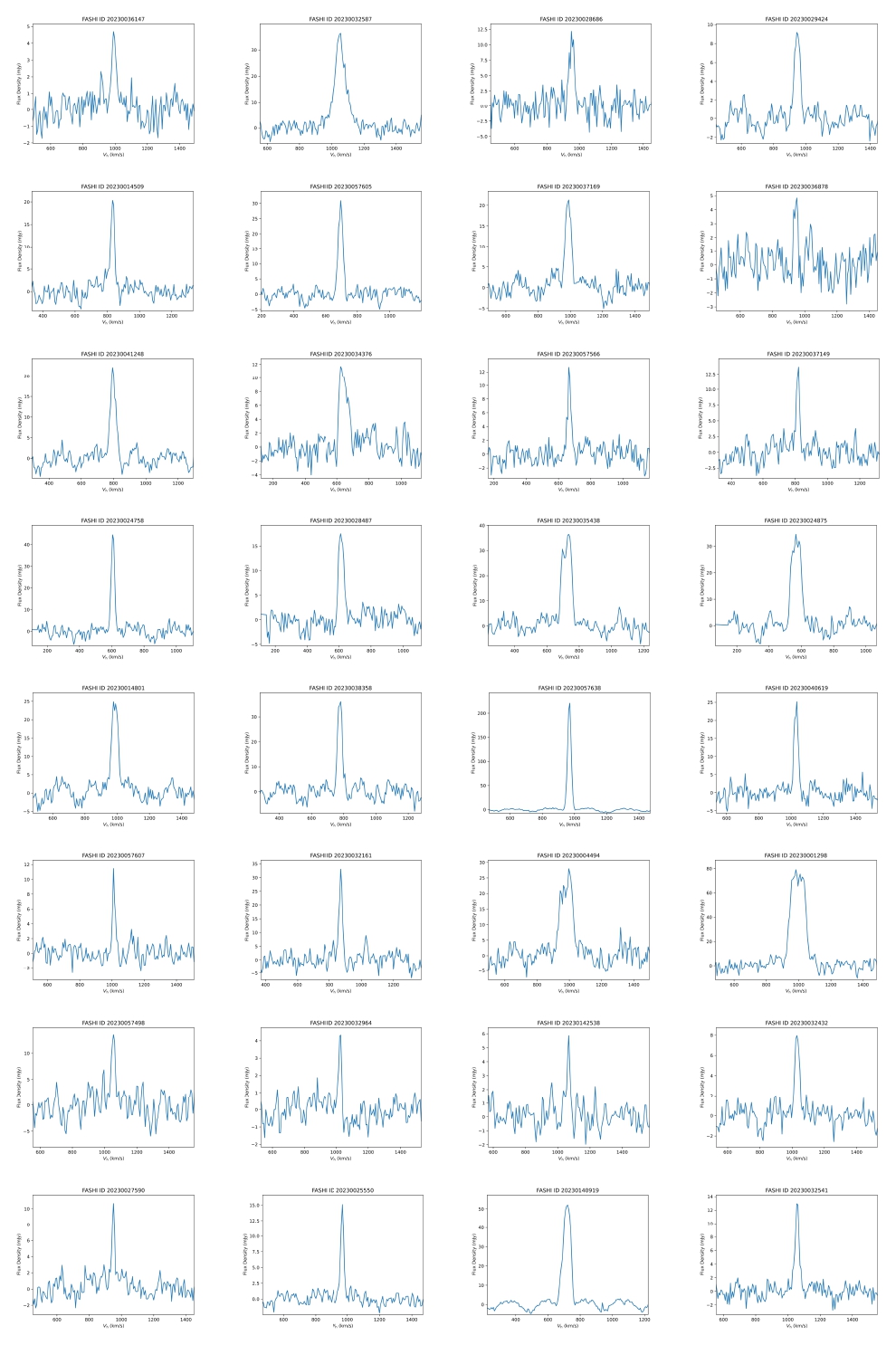}
\end{figure}

\addtocounter{figure}{-1}
\begin{figure}
\centering
\includegraphics[width=\linewidth]{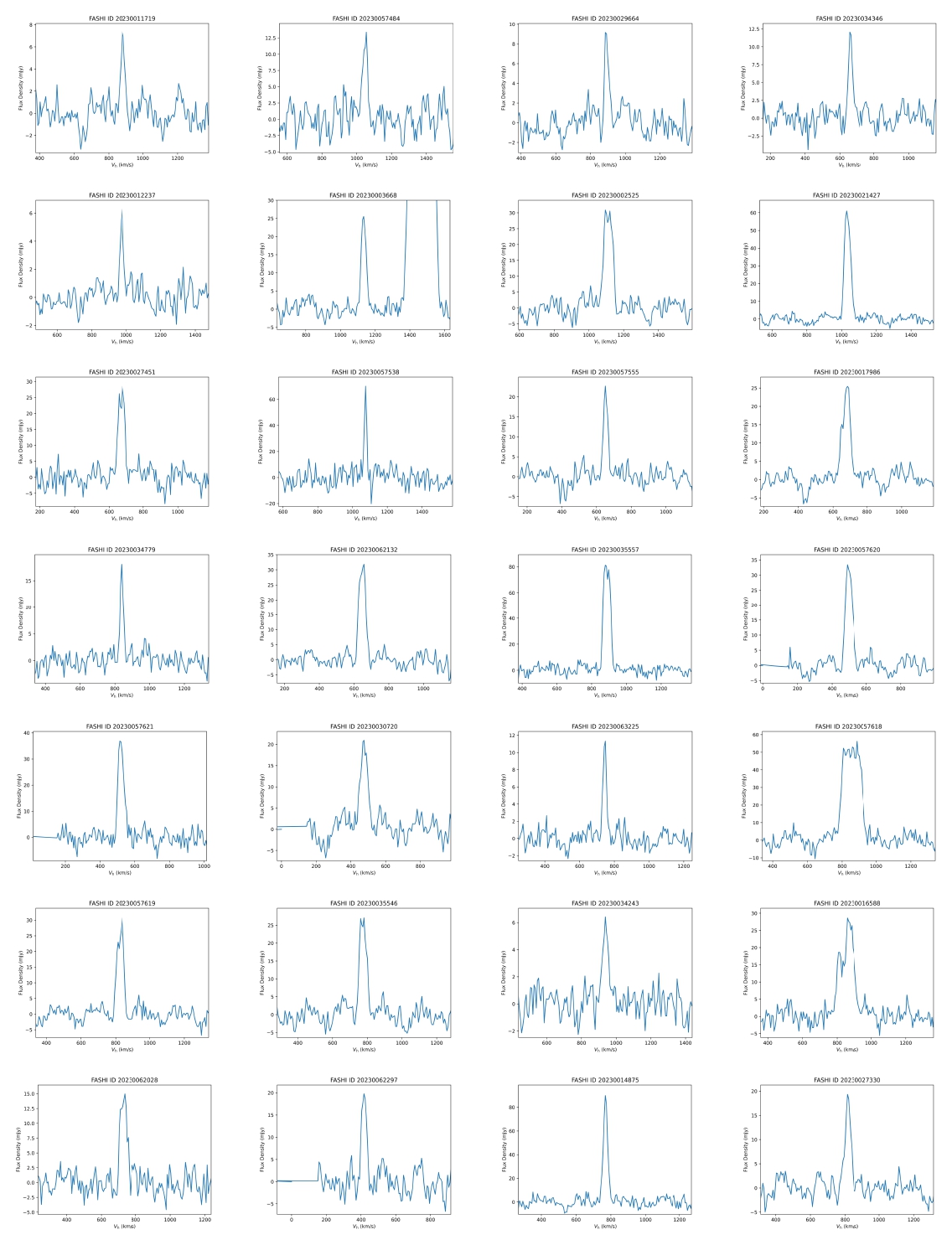}
\end{figure}

\label{lastpage}

\end{document}